\documentclass[aps,preprint,superscriptaddress]{revtex4-2}
\usepackage[utf8]{inputenc}
\usepackage[T1]{fontenc}
\usepackage{lmodern} 
\usepackage{graphicx,graphics,epsfig,epstopdf} 
\usepackage{dcolumn}
\usepackage{multirow}
\usepackage{amsmath,amssymb} 
\usepackage{enumitem}
\newcommand{\avg}[1]{\langle #1 \rangle}
\usepackage{fullpage}
\usepackage{tabularx,longtable,booktabs,diagbox}
\usepackage{natbib}
\usepackage[nottoc]{tocbibind}
\usepackage[english]{babel}
\usepackage[colorlinks=true,linkcolor=blue,citecolor=magenta,urlcolor=blue]{hyperref}  
\title{Bibliography Management: \texttt{natbib} Package}

\begin{document}

\title{Strong Gravitational Lensing by a Black Hole with a Global Monopole  in Kalb-Ramond Bumblebee Gravity}

\author{Bijendra Kumar Vishvakarma}
\email{bkv1043@gmail.com (BKV)}
\affiliation{Department of Physics, Banaras Hindu University,
Varanasi 221005, Uttar Pradesh, India}
\author{Shubham Kala}
\email{shubhamkala871@gmail.com (SK)}
\affiliation{The Institute of Mathematical Sciences, C.I.T. Campus, Taramani-600113, India}
     
\begin{abstract}
We investigate the strong gravitational lensing and shadow properties of the black hole in the context of bumblebee gravity, characterized by a global monopole charge $\kappa\eta^2$ and a Lorentz symmetry breaking parameter $\gamma$. We compute the deflection angles of light passing near the black hole in strong deflection limit, and estimate key lensing observables, including relativistic Einstein rings, absolute magnifications, image separations, and flux ratios, for astrophysical black holes. The black hole shadow is analyzed using the apparent angular size $\theta_{\rm Shadow} = 2\,\theta_{\infty}$ in the limiting photon orbit. Furthermore, we study the modification of the shadow structure in the presence of a radially infalling, optically thin accretion flow within a generalized framework. Our results indicate that both the global monopole charge and Lorentz-violating parameters significantly influence the photon sphere, lensing observables, and shadow morphology, potentially providing observational signatures for testing bumblebee gravity in the strong-field regime.  

\vspace{0.5em}
\end{abstract}

\maketitle

\newpage

\section{Introduction}
One of the most fascinating, intriguing, and extremely dense objects predicted by the general theory of relativity is the \emph{black hole} (BH)~\cite{Renn2005einstein,Schwarzschild1916,Birkhoff1923,Einstein1936,Darwin1959,Ohanian1987,Hawking1970,Penrose1965}. BHs are remarkably simple objects, as they are characterized by only a few parameters, such as mass and angular momentum~\cite{Kerr1963,Liebes1964,Weinberg1972,Ellis1973,Walsh1979,Senovilla1998}. Luminet \emph{et al.} investigated the shadow image of a spherical Schwarzschild BH surrounded by an accretion disk using null geodesics~\cite{Luminet1979}. The lens equation for weak and strong gravitational lensing (SGL) was initially derived by Schneider \emph{et al.}~\cite{Schneider1992} and \cite{Refsdal1994}. The fundamental physics of weak and SGL was discussed in detail by Narayan \emph{et al.}~\cite{Narayan1992}. Miyoshi \emph{et al.} confirmed the existence of a BH through high rotational velocities in the sub-parsec region of the galaxy NGC~4258~\cite{Miyoshi1995}. Astrophysical observations over the past decades indicate the existence of supermassive BHs~\cite{Frittelli2000,Begelman2003}. The Event Horizon Telescope (EHT) and LIGO--Virgo collaborations identified two supermassive BHs at the centers of our nearby galaxies, $SgrA^{*}$ and $M87^{*}$~\cite{Akiyama2019L1,Akiyama2019L2,Akiyama2019L3,Akiyama2019L4,Akiyama2019L5,Akiyama2019L6,Akiyama2019L12,Akiyama2019L17,Akiyama2019L25}. Astrophysicists and astronomers use BHs as natural laboratories to test various theories of gravity through both theoretical modeling and observational data. SGL probes the motion of photons in the vicinity of the photon region of a BH. Since gravity is not a force but a manifestation of spacetime curvature, light deviates from rectilinear propagation in strongly curved spacetime. The bending of light by massive objects, such as neutron stars or BHs, has been extensively confirmed and used to investigate weak and SGL in various BH spacetimes within general relativity~\cite{Walsh1979,Penrose1965,Kerr1963,Ellis1973,Senovilla1998}. The bending of light using the material medium approach has also been reported in the literature~\cite{Sen2010zzf,Roy2015hca,Roy2025hdw,Roy2025qmx}. In addition, despite astrophyscal significance light bending in lower-dimensional BH geometries has been studied to analyze the nature of lensing in such geometry~\cite{Kala2020viz,Kala2021ppi,Kala2025iri}. Massive objects responsible for bending light are known as lenses or deflectors. Therefore, the dynamics of photons in the strong gravitational field of BH spacetimes provide crucial information about BH properties. K.~S.~Virbhadra and collaborators explored the formation of an infinite set of relativistic images on both sides of the lens in SGL by spherically symmetric BHs in a series of seminal works~\cite{Virbhadra1998,Virbhadra2000,Virbhadra2002,virbhadra2008time,Virbhadra2009,Virbhadra2024compactness}. The exact gravitational lens equation valid in the strong-field regime, now known as the Virbhadra--Ellis lens equation, was formulated in Ref.~\cite{Virbhadra2000}. This equation has been widely applied to different BH spacetimes~\cite{Rabinowitz2004,Vazquez2004,Whisker2005,Gyulchev2007,Perlick2004,Perlick2007,Bin-nun2010,Sharif2011b,Horvath2011,Ding2011,Sharif2012,kocsis2013}.

Bozza extensively analyzed SGL in spherically symmetric spacetimes using the strong-deflection limit approach~\cite{Bozza2001,Bozza2002,Bozza2003,Bozza2005,Bozza2006strong,Bozza2006weakly,Bozza2007,Bozza2008,Bozza2010,Iyer2007,Kogan2008}. Gravitational lensing has been further investigated for various spherically symmetric BHs by many authors~\cite{Hsieh2021,Chagoya2021}. Chakraborty \emph{et al.} studied SGL in the presence of a Kalb--Ramond field~\cite{Chakraborty2017,Gonzalez2018,Zhang2018}. In this context, several authors have elaborated SGL in specific spacetime geometries~\cite{Ednaldo2024,Yang2023,DZYang2024,Shodikulov2025}. SGL by different BH spacetimes has also been discussed in Refs.~\cite{Bhadra2003,Nandi2006,Amore2007,Mukherjee2007,Ghosh2010,Sahu2012,Gyulchev2013,Wei2015,Guansheng2024,Saha2024,Pantig2025}.

The Bardeen regular BH is interpreted as a magnetic monopole solution arising from nonlinear electrodynamics~\cite{Bardeen1968,Dymnikova1992,Ayon1999,AyonBeato2000}. Several authors, including ourselves, have investigated gravitational lensing in this spacetime in the presence of a cloud of strings~\cite{Islam2024strong,Bkv2023,Bkv2025strong,AhmedKala2025}. Kala \emph{et al.} studied SGL by a non-minimally coupled Horndeski BH in a plasma medium~\cite{KalaSingh2025}. In addition to spacetime curvature, the topology of spacetime plays a significant role in gravitational lensing. Therefore, it is essential to analyze SGL induced by monopole defects. Barriola and Vilenkin investigated the gravitational field outside a monopole arising from the spontaneous breaking of global $O(3)$ symmetry~\cite{Barriola1989}. A global monopole does not exert a gravitational force on non-relativistic matter but instead produces a deficit solid angle in spacetime. Lee and Nair studied BHs with magnetic monopoles and showed that, for a limiting value of the Higgs-field vacuum expectation value, the monopole becomes a BH~\cite{LeeNair1992}. Global monopoles in de~Sitter and anti--de~Sitter spacetimes were studied in Refs.~\cite{LiHao2002,BBH2003}. Brihaye \emph{et al.} analyzed global monopoles and BHs with monopole hair in the Einstein--Goldstone model~\cite{Brihaye2005}. Topological defects formed during phase transitions in the early universe contribute to gravitational lensing phenomena. One such defect is the global monopole, which may have been produced during early-universe phase transitions~\cite{cheng2003strong,Cheng2010,Man2015,Sharif2015}. Nucamendi \emph{et al.} analyzed the motion of test particles and computed frequency shifts in global monopole spacetimes~\cite{Nucamendi2000,Nucamendi2001}. Soares \emph{et al.} studied SGL in a topologically charged Eddington-inspired Born--Infeld BH~\cite{Soares2023}. Lan \emph{et al.} investigated the combined effects of electric charge and a global monopole on SGL~\cite{Lan2025}.

Beyond general relativity and the Standard Model, considerable interest has been devoted to BH spacetimes with Lorentz symmetry violation, motivated by possible signatures from string theory~\cite{Kostelecky1989,Esposito2011}. Such considerations lead to Lorentz symmetry breaking, often modeled using a bumblebee self-interacting vector field~\cite{Kostelecky2004}. Lorentz symmetry breaking has also been explored for rotating BHs~\cite{Jha2022,Universe2023,DO2024,Sucu2025}. Ara\'ujo Filho \emph{et al.} analyzed SGL by a Lorentz-violating BH~\cite{Nascimento2024}. Rukkiya \emph{et al.} and others studied SGL and BH shadows in the presence of a Kalb--Ramond field and plasma~\cite{RukkiyyaSini2025,Xu2025KRShadow,SucuSakalli2025KRPhoton,Shodikulov2025KRDM}.
Therefore, it is natural to investigate SGL by a global monopole in Ricci-coupled Kalb--Ramond bumblebee gravity. Gralla \emph{et al.} discussed BH shadows in optically and geometrically thin accretion disks using lensing rings~\cite{Gralla2019,Vincent2022Photon}. Aslam \emph{et al.} analyzed the shadow and observables of a dyonic BH with a global monopole under different accretion profiles in the presence of perfect fluid matter~\cite{AslamSaleem2024}.
\\
This paper is structured as follows. In Section~\ref{section2}, we briefly review the spacetime geometry and the effective potential, and describe the formalism of SGL in a spherically symmetric spacetime. In Section~\ref{section3}, we study the strong deflection limit coefficients. Section~\ref{section4} is devoted to the analysis of magnification and relativistic Einstein rings. In Section~\ref{section5}, we analyze the shadow properties of two astrophysical BHs, Sgr~A$^{*}$ and M87$^{*}$, located in nearby galaxies. The impact of a radially infalling accretion gas on the BH shadow is investigated in Section~\ref{section6}. Finally, we summarize our conclusions in Section~\ref{section7}. Throughout this paper, we work in natural units with $G = \hbar = c = 1$, unless stated otherwise.

\section{SGL by a BH with a Global Monopole  in a Ricci-coupled Kalb-Ramond Bumblebee Gravity} \label{section2}

An extension of Einstein-bumblebee gravity has been obtained  by Fernando M. Belchior et al. in the paper~\cite{Belchior2025global}. The basic action that coupled the Ricci tensor to Kalb Ramond field is given by
\begin{equation} 
    \mathcal{A}=\int d^{4}x \, \sqrt{-g}
\left[ \frac{1}{2 k }\left( R+\epsilon\ B^{\mu\lambda } B^{\nu}_{\lambda }R_{\mu \nu}\right) -\frac{1}{12}H_{\lambda \mu \nu}H^{\lambda \mu \nu}-V\left(B_{\mu \nu}B^{\mu \nu}\pm \ b^2\right)+ \mathcal{L}_{m} \right]\
\label{1}
\end{equation}
Where $g=$ is determinant of metric tensor $g_{\mu\nu}$ and $k=8\ \pi\ G_N$,$G_N$ is being the Universal gravitational constant. The coupling constant $\epsilon$ coupled the Kalb Ramond field $B_{\mu\nu}$ with Ricci tensor $R_{\mu\nu}$. Where KR field defined through third rank antisymmetric tensor $H_{\lambda \mu \nu}=\partial_{\lambda }B_{\mu\nu}+\partial_{\mu }B_{\nu\lambda}+\partial_{\nu }B_{\lambda\mu}$. The Lorentz symmetry breaking has non zero vacuum expectation value $\avg {B_{\mu\nu}}=b_{\mu\nu}$. The potential term have non zero value defined in terms of dual tensor field$\avg{\tilde B_{\mu\nu}}=\frac{1}{2}\avg {\epsilon_{\mu\nu \alpha \beta}{B^{\alpha\beta}}}=\frac{1}{2}\avg {\epsilon_{\mu\nu \alpha \beta}{b^{\alpha\beta}}}$. The Levi-Civita tensor $\epsilon_{\mu\nu \alpha \beta}$ is a total antisymmetric tensor. We obtain gravitational equation by varying action {(\ref{1})} with respect to metric tensor $g^{\mu\nu}$.
\begin{equation} \label{eq2}
    G_{\mu\nu}= \ k \left(T^{Matter}_{\mu\nu}+\ T^{KR}_{\mu\nu} \right)+T^{non-minimal}_{\mu\nu}
\end{equation}
Where $G_{\mu\nu}=R_{\mu\nu}-\frac{1}{2}\ g_{\mu\nu}\ R$ is Einstein tensor, ~$R$~ Ricci scalar and right terms are given by 
\begin{eqnarray}
T^{Matter}_{\mu\nu}=-\frac{2}{\sqrt{-g}}\frac{\delta\left(\sqrt{-g}\ \mathcal{L_{m}}\right)}{\delta g^{\mu \nu}}\\
T^{KR}_{\mu\nu}=-\frac{1}{2}H_{\alpha\beta\mu}H^{\alpha\beta}_{\nu}-\frac{1}{12}g_{\mu\nu} \ H_{\lambda\alpha\beta}H^{\lambda\alpha\beta}-g_{\mu\nu}V+4B_{\alpha\mu}B^{\alpha}_{\nu}V^{'}\\
T^{non-milimal}_{\mu\nu}=\frac{\epsilon}{k}\Bigg[\frac{1}{2}g_{\mu\nu}B^{\alpha\lambda}B^{\beta}_{\lambda}R_{\alpha\beta}-B_{\alpha\mu}B_{\beta\nu}R^{\alpha\beta}-B_{\alpha\beta}B^{\alpha}_{\nu}R^{\beta}_{\nu}-B_{\alpha\beta}B^{\alpha}_{\nu}R^{\beta}_{\mu}+\\ 
\frac{1}{2}\nabla^{\alpha}\nabla_{\mu}B^{\beta}_{\nu}R_{\alpha}+\frac{1}{2}\nabla^{\alpha}\nabla_{\nu}\ B^{\beta}_{\mu}\ R_{\alpha\beta}-\frac{1}{2} \Box B^{\alpha}_{\mu} B^{\alpha}_{\nu}-\frac{1}{2}g_{\mu\nu} \nabla_{\alpha}\nabla_{\beta}B^{\alpha\lambda}B^{\beta}_{\lambda}\Bigg]
\end{eqnarray}
where $'$ for differentiation with respect to argument of Lorentz violating potential.
Also we get following equation by varying {(\ref{1})} with respect to field strength $B_{\mu \nu}$
\begin{equation}
    \nabla_{\lambda}H^{\lambda\mu\nu}=4V^{'}B^{\mu\nu}+\frac{\epsilon}{k}\Bigg[B^{^\mu\lambda} R^{\nu}_{\lambda}-B^{^\nu\lambda} R^{\mu}_{\lambda}\Bigg]
\end{equation}
The matter stress tensor associated with global monopole described by Lagrangian density
\begin{equation}
    \mathcal{L_{m}}=\frac{1}{2}\ \partial{\mu}\ \phi^p \ \partial{^\mu}\ \phi^p-\frac{\xi}{4}\  \left(\phi^p \ \phi^p-\eta^2\right)^2
\end{equation}
Here $p$ labels scalar field $\phi^p=\phi^1, \phi^2, \phi^3$ and $\xi ,v $ for coupling constant and energy scale respectively . Outside  core of global monopole is characterize by following ansatz $\phi^p=v\frac{x^p}{r},\ x^p x^p=r^2$. Consider spherically symmetric solution of non zero value of KR field
\begin{equation}
ds^2=-f(r)\ dt^2+ \frac{1}{f(r)}\ dr^2+ r^2 d\Omega_2 ,
\label{sph}
\end{equation}
where, $d\Omega_2=(d\theta^2+\sin^2\theta d\phi^2)$ is 2-sphere.
The diagonal non zero components of energy momentum tensor of global monopole\cite{Rhie1991,Dadhich1998} and Lorentz violating Kalb-Ramond field\cite{Lessa2020,Yang2023} given by 
\begin{align}
T^{\text{Matter}}_{\mu\nu}
&= \mathrm{diag}\left(\frac{f(r)v^2}{r^2},\frac{v^2}{f(r)r^2},0,0\right),
\label{eq10} \\
b_{\mu\nu}
&= b_{01}=-b_{10}=\frac{b}{\sqrt{2}} .
\label{eq11}
\end{align}
Using equations \eqref{eq10}, \eqref{eq11} and \eqref{eq2}, we have solution of the Einstein field equation expressed as follows
\begin{equation}
ds^2=-\left(\frac{1-\kappa\eta^2}{1-\gamma}-\frac{2M}{r}\right)dt^2+ \frac{1}{\left(\frac{1-\kappa\eta^2}{1-\gamma}-\frac{2M}{r}\right)}dr^2+ r^2 (d\theta^2+\sin^2\theta d\phi^2).
\label{eqn:12}
\end{equation} 
Here, $M$ denotes the mass of black hole. The parameter $\kappa\eta^{2}$ represents the global monopole term, while $\gamma = \frac{\epsilon b^{2}}{2}$ characterizes Lorentz symmetry breaking and is commonly referred to as the Lorentz-violating (LV) parameter~\cite{Belchior2025global}.     
The Event horizon of the BH metric is given by the roots of $ f(r)=0$
  and is located at, 
  \begin{align*}
     r_{h}&=\frac{2M(1-\gamma)}{(1-\kappa  \eta^2)}.
\end{align*}
 \begin{figure*}[ht]
\begin{tabular}{c c }
\includegraphics[width=0.45\linewidth, height=0.3\textheight]{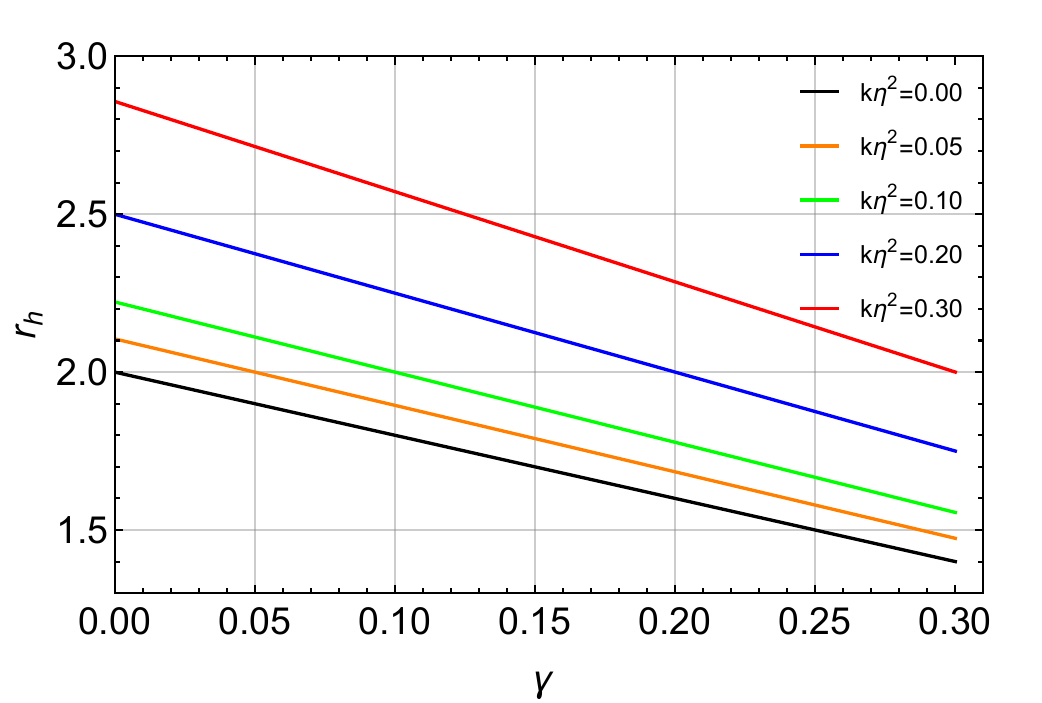}
\includegraphics[width=0.45\linewidth, height=0.30\textheight]{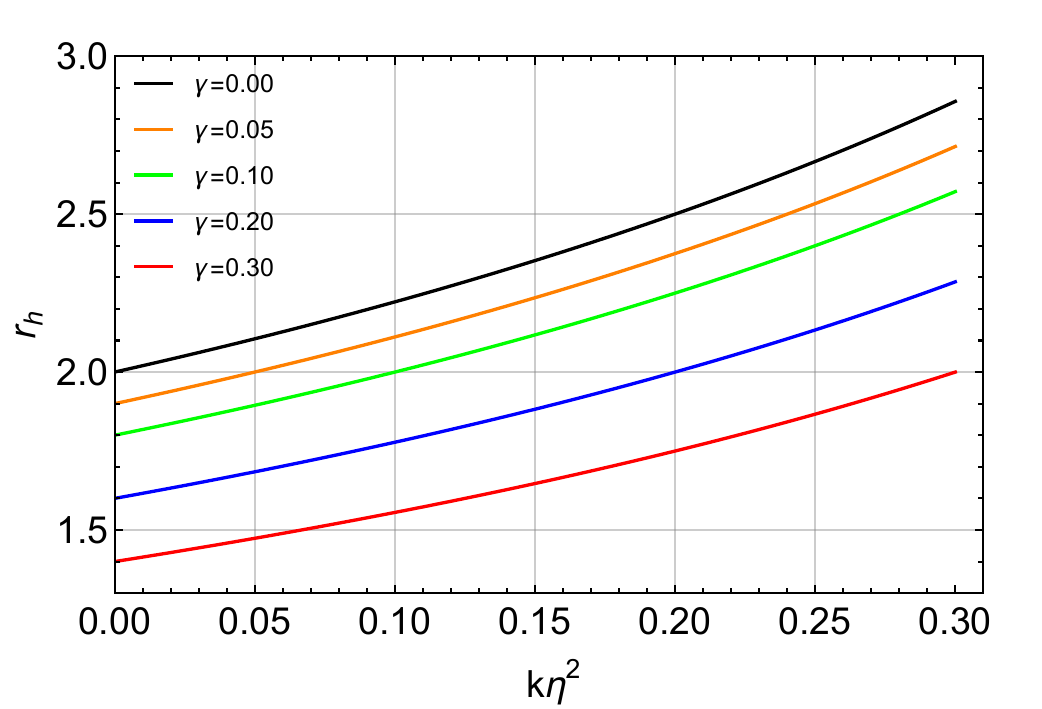}\\
\end{tabular}
\caption{The plot of event horizon, versus BH parameter  $\gamma$ (left) and $\kappa\eta^{2}$ (right)}.
\label{Fig:1}
\end{figure*}
The event horizon radius increases with the global monopole parameter and decreases with the LV parameter, as shown in Fig.~(\ref{Fig:1}). Consequently, the gravitational field of the BH weakens due to the global monopole parameter and strengthens due to the LV parameter.
In units of $2M=1$, with the dimensionless radial coordinate $x=r/2M$,$T=t/2M$ the metric functions of the BH spacetime equation~(\ref{eqn:12}) with the global monopole term are expressed as
\begin{align}
A(x) &= B(x)^{-1} = 1 - P - \frac{1}{x}, \nonumber\\
C(x) &= x^{2}, \qquad
P = \frac{\kappa\eta^{2} - \gamma}{1 - \gamma}.
\end{align}
The corresponding line element takes the form
\begin{equation}
ds^{2} = -A(x)\, dT^{2} + B(x)\, dx^{2} + C(x)\left(d\theta^{2} + \sin^{2}\theta\, d\phi^{2}\right).
\end{equation}

\subsection{Effective potential and impact parameter}
As the null geodesic passes  near a BH, it deviates from its original path due to strong gravitational field. In strong deflection limit, we consider light rays propagating close to the spherical photon sphere in the equatorial plane~${\theta=\frac{\pi}{2}}$. The effective potential is given by,
\begin{eqnarray}
\frac{dx}{d\tau}+\mathcal{V}_{eff}(x)=1,
 \end{eqnarray} 
where ${\tau}$ is affine parameter, and $\mathcal{V}_{eff}$ is the effective potential of BH,
\begin{eqnarray}
\mathcal{V}_{eff}(x)=\frac{u^2}{x^3}\left[\frac{x(1-\kappa\eta^2)}{(1-\gamma)}-1\right]
 \end{eqnarray}
 The impact parameter is defined as ${u_{phs}}=L/E$. we plot the effective potential for fixed values of $\gamma = 0.1$ and $\kappa\eta^{2} = 0.1$ (left panel). In the right panel, we fix $\gamma = 0.1$ and $u = 2.608$. The case corresponding to equal values of the BH parameters reduces to Schwarzschild black hole, denoted by the subscript (S). The presence of a single maximum in the effective potential confirms the existence of an unstable circular orbit. The unstable spherical photon orbit radius is determined with the condition where deflection angle diverges and  obtained by using the following expressions together,
\begin{align*}
 \frac{dx}{d\tau}=0, ~~~ &\frac{d^2x}{d\tau^2}=0,~~~
\frac{d^2\mathcal{V}_{eff}(x)}{d\tau^2}<0&.
\end{align*}  

\begin{figure*}[ht]
    \begin{tabular}{c c }
    \includegraphics[width=0.45\linewidth,height=0.45\linewidth]{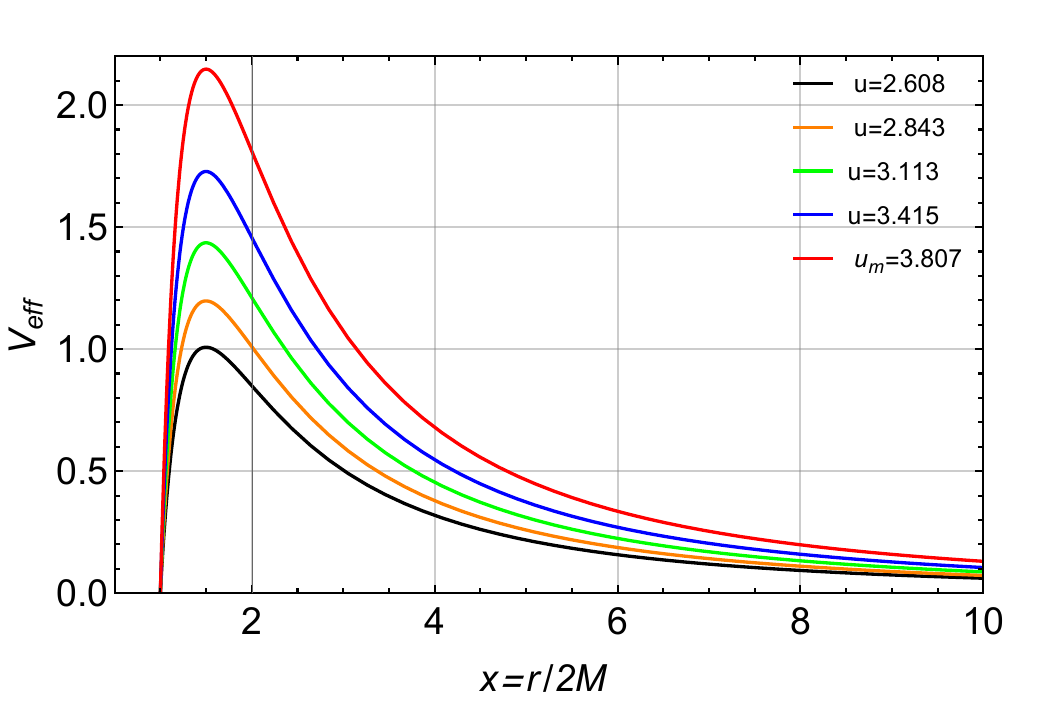}
    \includegraphics[width=0.45\linewidth,height=0.45\linewidth]{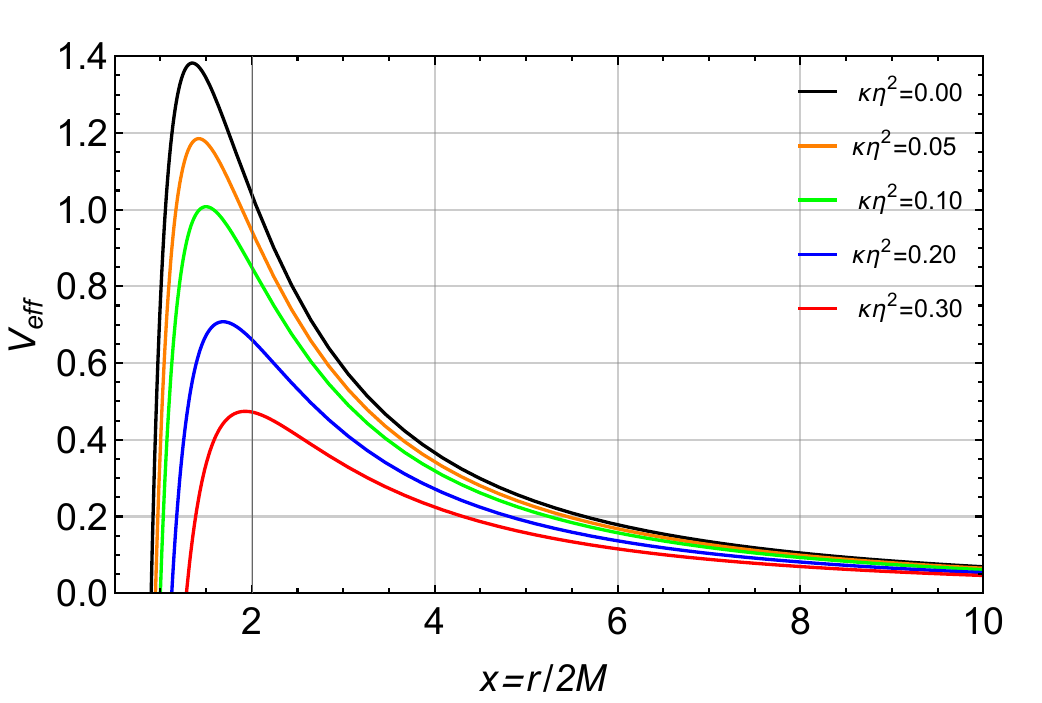}\\
    \end{tabular}
    \caption{Effective potential at fixed value of $\gamma=0.1,\kappa\eta^2=0.1$ (left). In right plot $\gamma=0.1,u=2.608$. Equal values of BH parameter correspond to Schwarzschild BH, denoted by subscript (S).}
    \label{Fig:2}
\end{figure*}
\begin{figure*}[ht]
\begin{tabular}{c c }
\includegraphics[width=0.45\linewidth, height=0.30\textheight]{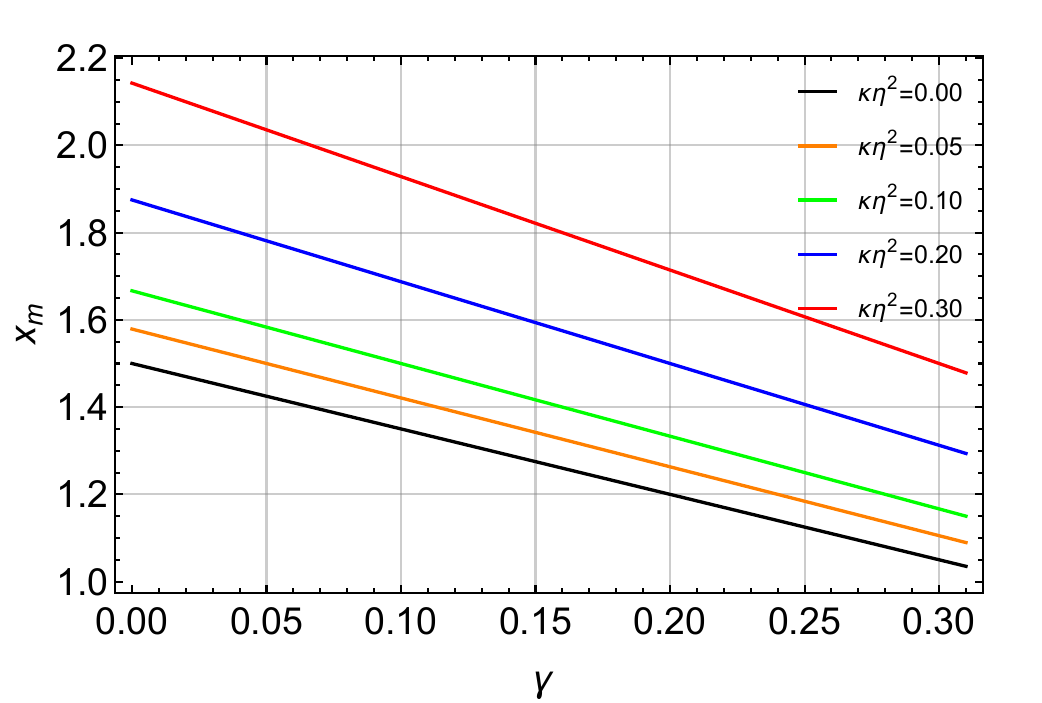}
\includegraphics[width=0.45\linewidth, height=0.30\textheight]{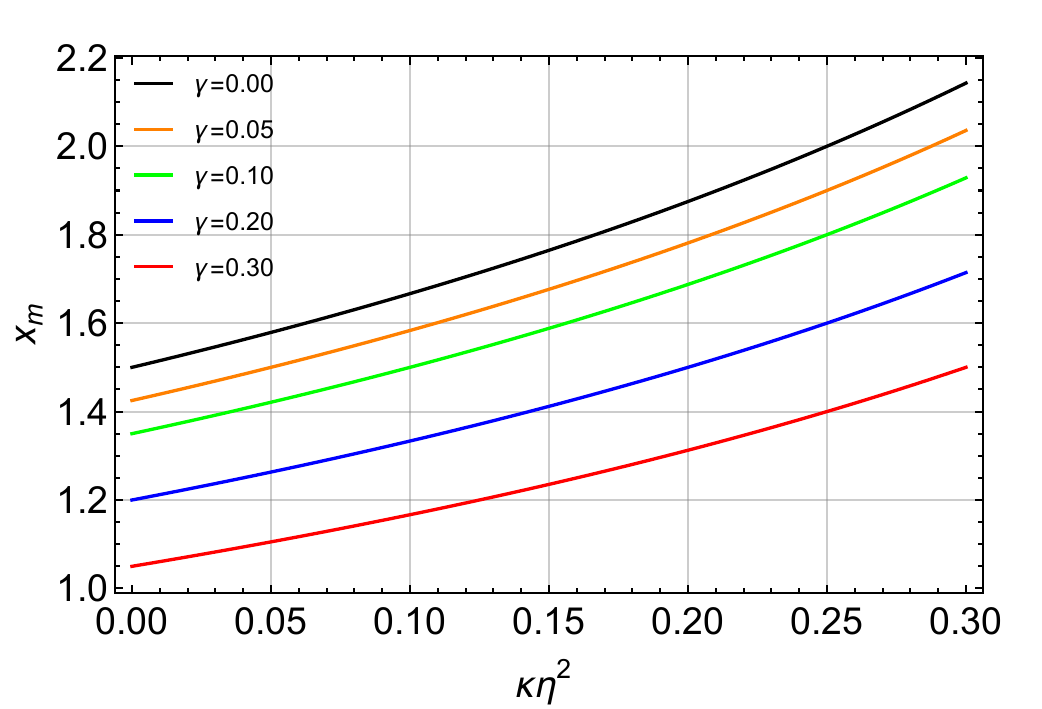}\\
\end{tabular}
\caption{The plot of photon radius,versus BH parameter  $\gamma$ (left) and $\kappa\eta^{2}$ respectively (right)}.
\label{Fig:3}
\end{figure*}
\begin{figure*}[ht]
\begin{centering}
\includegraphics[width=0.48\linewidth, height=0.35\textheight]{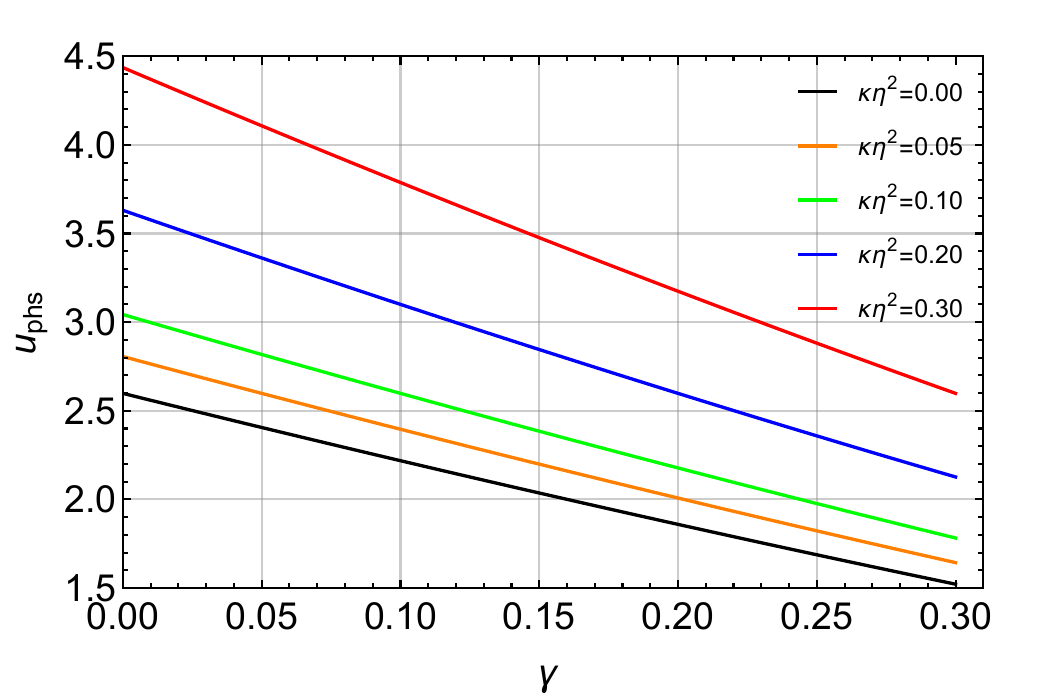}
\includegraphics[width=0.48\linewidth, height=0.35\textheight]{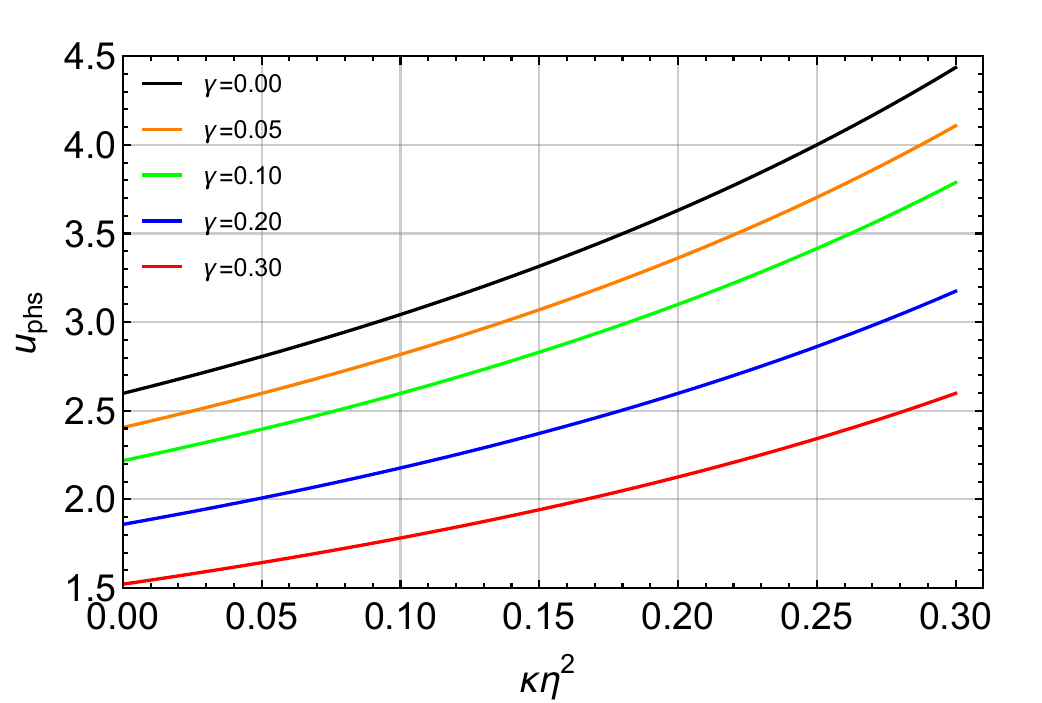}
\caption{Plot of impact parameter for versus $\gamma$ \ (left) and $\kappa\eta^2$ \ (right) respectively.}
\label{Fig:4}
\end{centering}
\end{figure*}

These spherical photon radii forms the time like hypersurfaces.
Photon radius is given by following condition \cite{Bozza2002,Tsukamoto2017}-
\begin{eqnarray}
    \frac{A'}{A}=\frac{C'}{C}, \ &&
    x_{p}=\frac{3(1-\gamma)}{2(1-\kappa\eta^2)}
\end{eqnarray}
The Fig.~(\ref{Fig:3}) and Fig.~(\ref{Fig:4}) shows variation of photon radius and impact parameter with $\gamma$ and $\kappa \eta^2$. The both radius increases with global charge parameter and decrease with $\gamma$. Hence both parameter of BH have opposite effects on light rays. This variation have a significant effects on deflection angle.

\newpage
\section{Strong deflection limit coefficients}  \label{section3}
\begin{figure*}[t]
  \centering
  \includegraphics[width=0.75\textwidth]{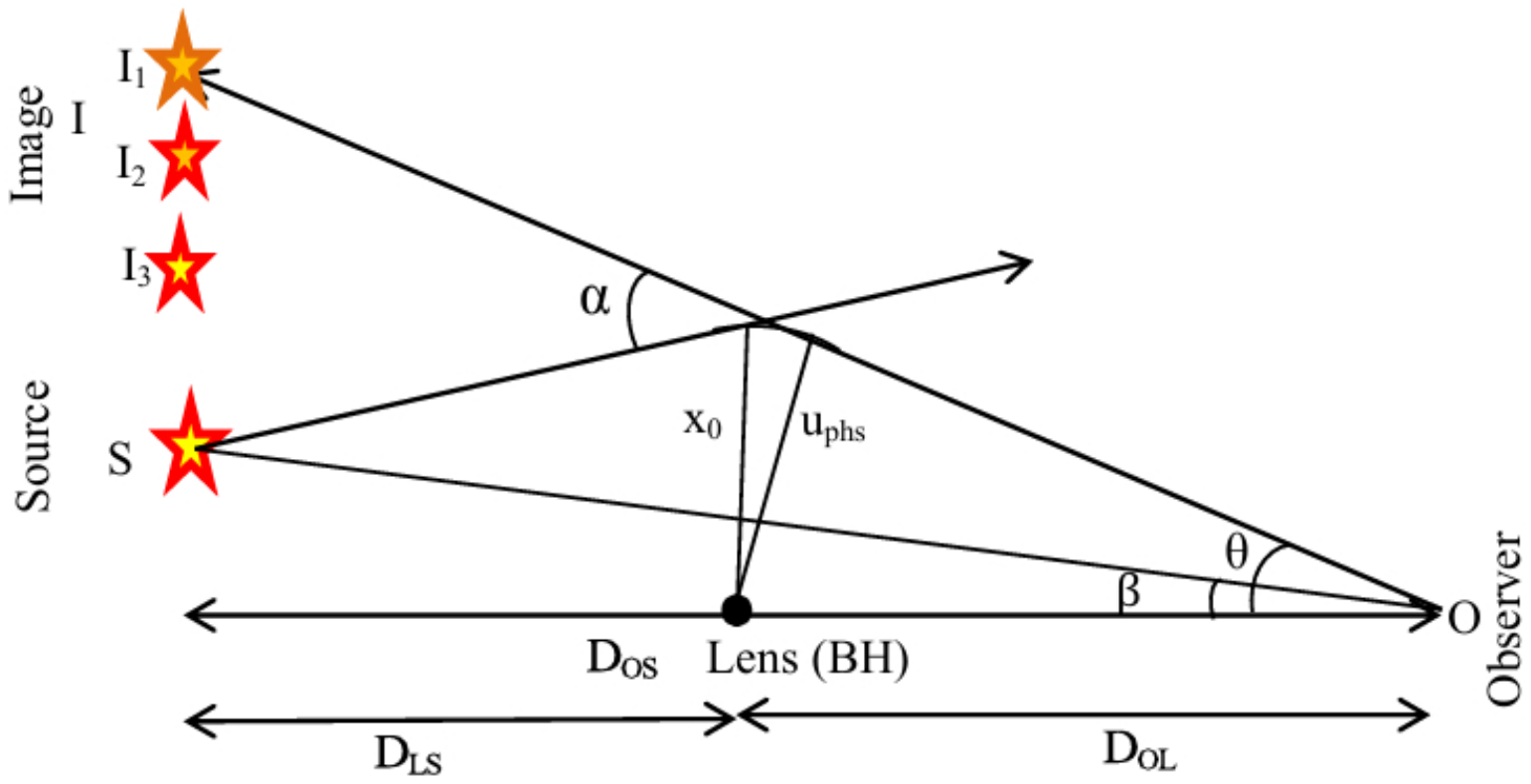}
  \caption{Schematic diagram of strong gravitational lensing.}
  \label{Fig:5}
\end{figure*}

 In this section, we analyze the deflection angle of null geodesics and coefficients of strong deflection limit. Consider the source and observer  situated at large distances in flat spacetime. Light rays travel towards lens (BH) from the source in strong gravity of BH. Consequently photons have deflected and detected by observer (O). This journey of light have demonstrated in Fig.~(\ref{Fig:5}).

Fig.~(\ref{Fig:5}) shows the schematic diagram of SGL.Among infinite images only handful of relativistic images are shown in one side. The observer can see an infinite images of the source. We have only show few of them and an outermost image at  angular position $\theta$. These images has been formed by the rays suffered through deflection angle $\alpha$. Distances $D_{LS},D_{OL},D_{OS}$ are between Lens-Source,Observer-Lens, Observer-Source respectively. The arrow with head showing the tangents on the source ray and at the observer for images that formed due to SGL. The $x_{0}$ denotes closest approach of light rays to the BH.
Virbhadra et al.~\cite{Virbhadra1998} described the  angular shifts of photon geodesics in terms of the function of distance from center in spherically symmetric BH spacetime as given below: 
\begin{eqnarray}
\frac{d\phi}{dx}=\frac{\sqrt{B}}{\sqrt{C}\sqrt{\frac{C}{C_{0}}(\frac{A_{0}}{A})-1}};
\label{eqn:18}
\end{eqnarray}
Subscript $0$ represents that the metric functions are calculated at minimum approach $x_{0}$ from the center of the lens. Also special form of equation~\eqref{eqn:18} was used by S. Weinberg (see chapter $(8.4-8.5)$) in~\cite{Weinberg1972}.The deflection angle enhances when distance of closest approaches decreases then going to diverge at $x_{0}$ equals to $x_{m}$ photon radius~\cite{Virbhadra2000,Claudel2001geometry,Bozza2002}. The deflection angle of light rays that is deviates by its original trajectory due to strong gravitational field of lens (deflector) encoded by Virbhadra et al.~\cite{Virbhadra1998} as follows equation containing integral of equation~(\ref{eqn:18}) for motion of photon,
\begin{equation}
    \alpha_{D}(x_{0})=I(x_{0})-\pi;
\end{equation}
where,
\begin{equation}
    I(x_{0})=2\int_{x_{0}}^{\infty}\frac{d\phi}{dr}dx=\int_{x_{0}}^{\infty}\frac{\sqrt{B}}{\sqrt{C}\sqrt{\frac{C}{C_{0}}(\frac{A_{0}}{A})-1}}dx.
\end{equation}
Using change of variable $z=\frac{A-A_{0}}{1-A_{0}}$ the above integral can be recast as \cite{Eiroa2011,Zhang2017,Tsukamoto2017,Tsukamoto2021}.
This integral describes the photon dynamics analytically in  BH spacetime as follows,

 \begin{equation}
      I(x_{0})=\frac{2x^{2}}{x_{0}}\int_{0}^{1}R(z,x_{0})f(z,x_{0})dz,\\
 \end{equation}
      where,

     \begin{align*}
     R(z,x_{0})=\frac{2\sqrt{AB}}{C A'}(1-A_{0})\sqrt{C_{0}},\\
      f(z,r_{0})=\frac{1}{\sqrt{A_{0}-\frac{C_{0}}{C}[(1-A_{0})z]}}.
 \end{align*}

Expanding $ f(z,x_{0})$ by Taylor's expansion upto $z^2$ which yields \cite{Eiroa2011},
\begin{equation}
    f(z,x_{0})=\frac{1}{\sqrt{\alpha\ z+\beta\ z^2}},
\end{equation}
where $\alpha$ and $\beta$ are given by the following expressions,
\begin{align*}
    \alpha=\frac{1-A_{0}}{C_{0}A_{0}'}(C_{0}'A_{0}-A_{0}'C_{0}),\\
    \beta=\frac{(1-A_{0})^{2}}{C_{0}^{2}A_{0}^{3}}[2C_{0}C_{0}'A_{0}'^{2}-A_{0}A_{0}''C_{0}'+A_{0}A_{0}'(C_{0}''C_{0}-2C_{0}'^{2})].
\end{align*}
 
\begin{figure*}[ht]
\begin{tabular}{c c}
\includegraphics[width=0.45\linewidth, height=0.35\textheight]{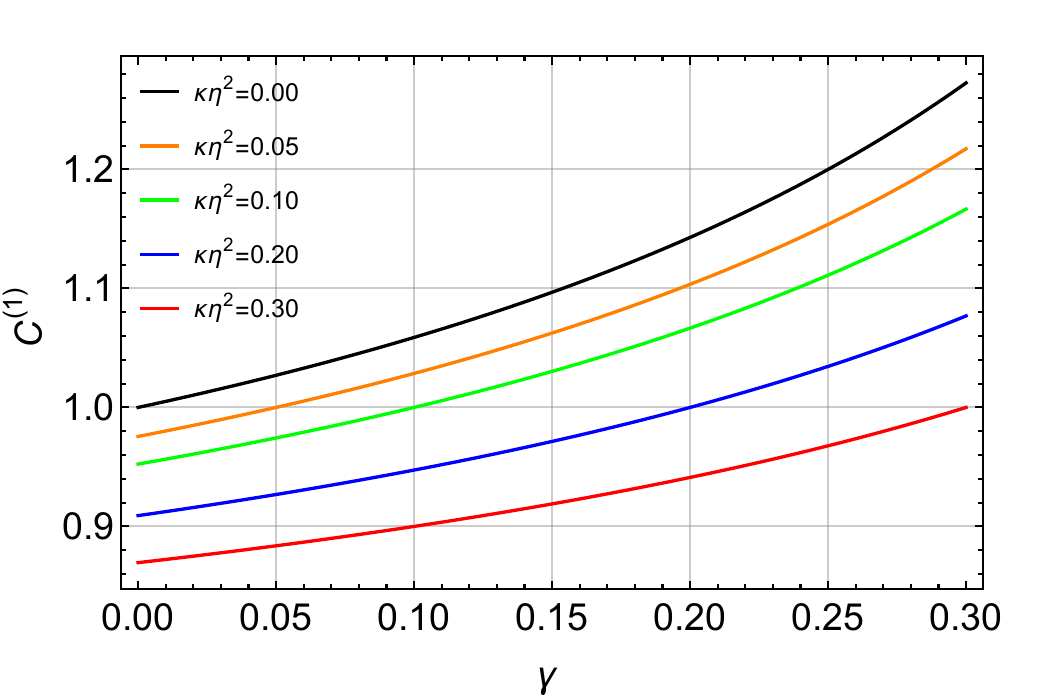}
\includegraphics[width=0.45\linewidth, height=0.35\textheight]{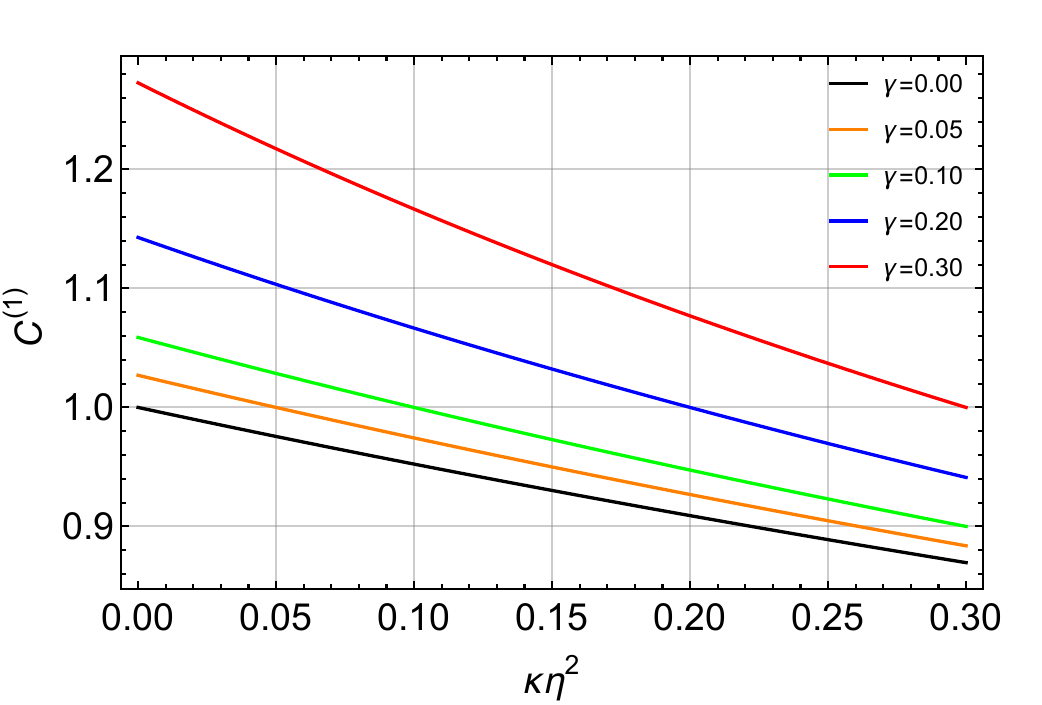}\\
\end{tabular}
\caption{First deflection coefficient for fixed value of $\gamma$ and $\kappa\eta^2 $ (right) respectively.}
\label{Fig:6}
\end{figure*}  
The strong lensing coefficients are given by,
\begin{equation}
    C^{(1)}=\frac{R(0,x_{ph})}{2\sqrt{\beta}},
\end{equation}
where $x_{m}=x_{ph}$ is largest photon radius and 
\begin{equation}
    C^{(2)}==-\pi+C_{R}+C^{(1)}\ln2\ \delta,
\end{equation}
where, $\delta=\frac{2\ \beta }{A(x_{m})}$ \& $C_{R}=I(x_{m})$\ where $ P=\frac{\kappa\eta^2-\gamma}{1-\gamma}$.

In terms of BH parameter strong deflection limit coefficient are given by 
\begin{eqnarray}
    C^{(1)}=\frac{2}{2+P}, \ \\
     C^{(2)}=-\pi+\frac{4\ log\Bigg[\frac{6\ \sqrt{1-P}}{\sqrt{3-12P}+3\sqrt{1-P}}\Bigg]}{\sqrt{1-P}}-2\ \frac{log{\Bigg[\frac{3\ (2+P)^2}{2\ (1-P)^2}\Bigg]}}{2+P}.
\end{eqnarray}

\begin{figure*}[ht]
\begin{tabular}{c c}
\includegraphics[width=0.45\linewidth, height=0.35\textheight]{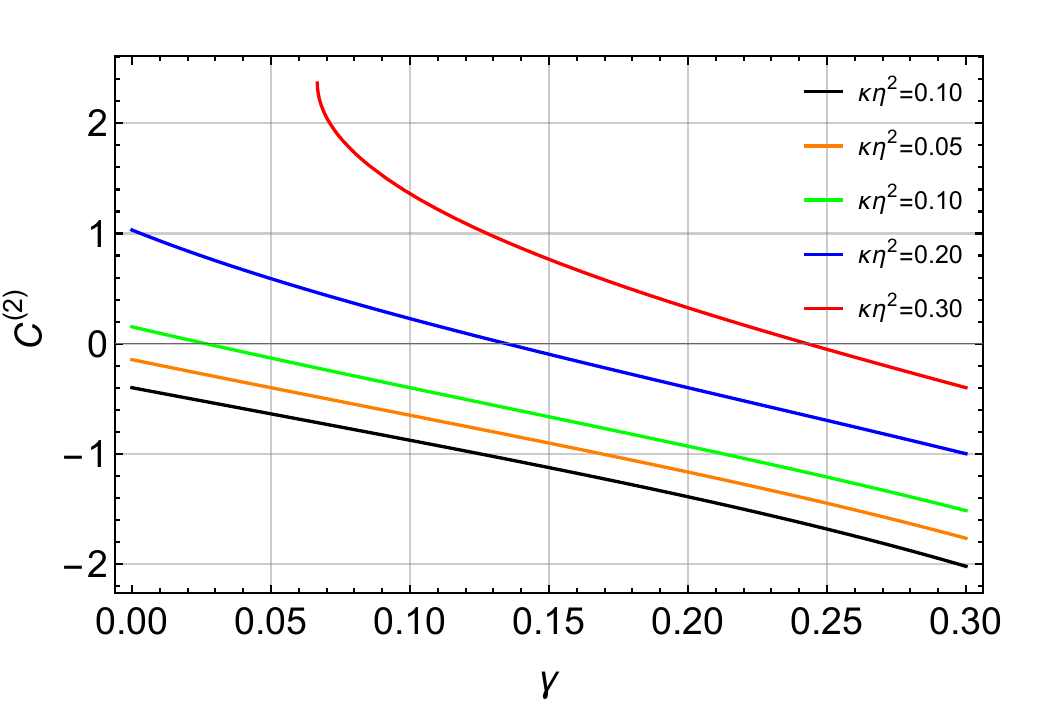}
\includegraphics[width=0.45\linewidth, height=0.35\textheight]{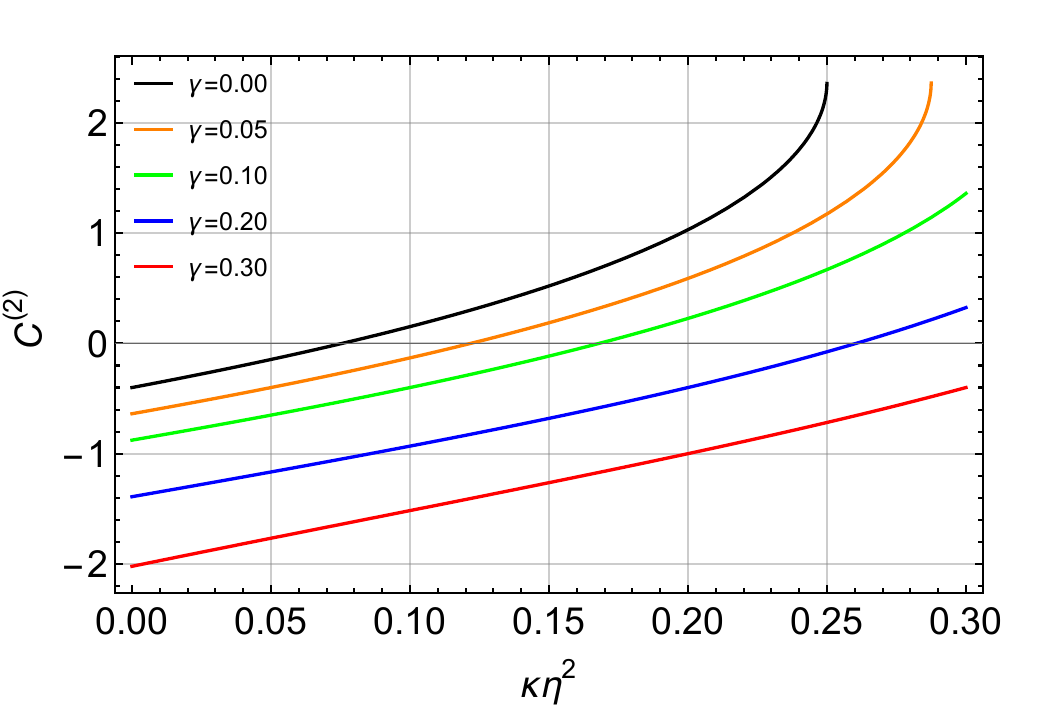}\\
 \end{tabular}
\caption{Second deflection coefficient for fixed value of $\gamma$ and $\kappa\eta^2 $ (right) respectively.}
\label{Fig:7}
\end{figure*}
\begin{figure*}[ht]
\begin{tabular}{c c }
\includegraphics[width=0.45\linewidth, height=0.35\textheight]{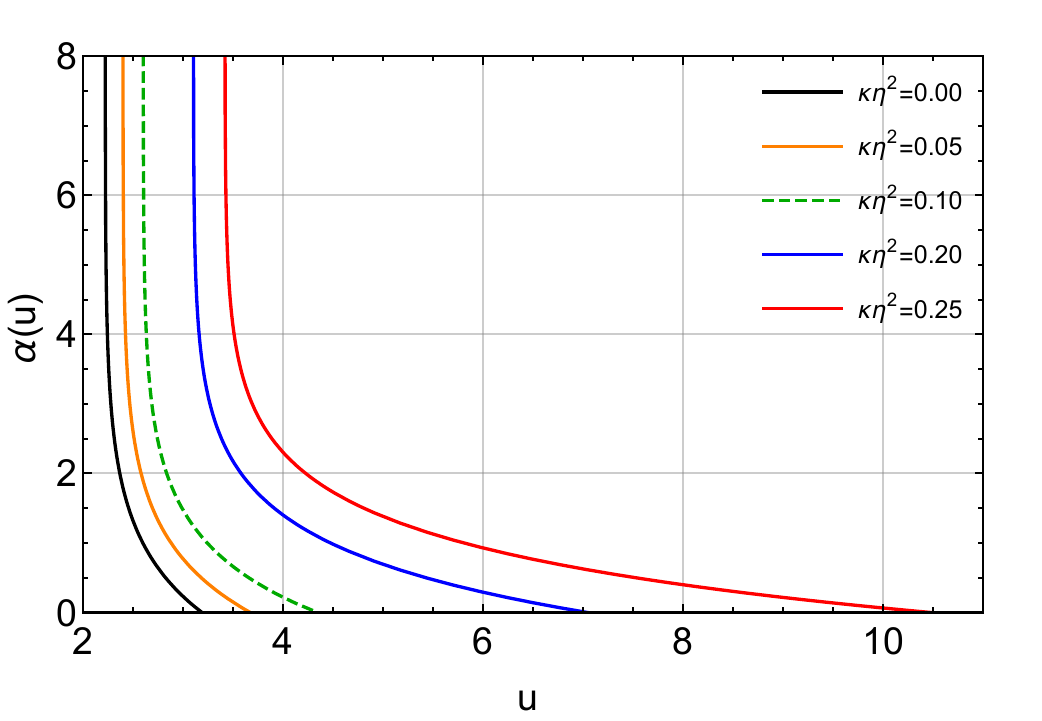}
\includegraphics[width=0.45\linewidth, height=0.35\textheight]{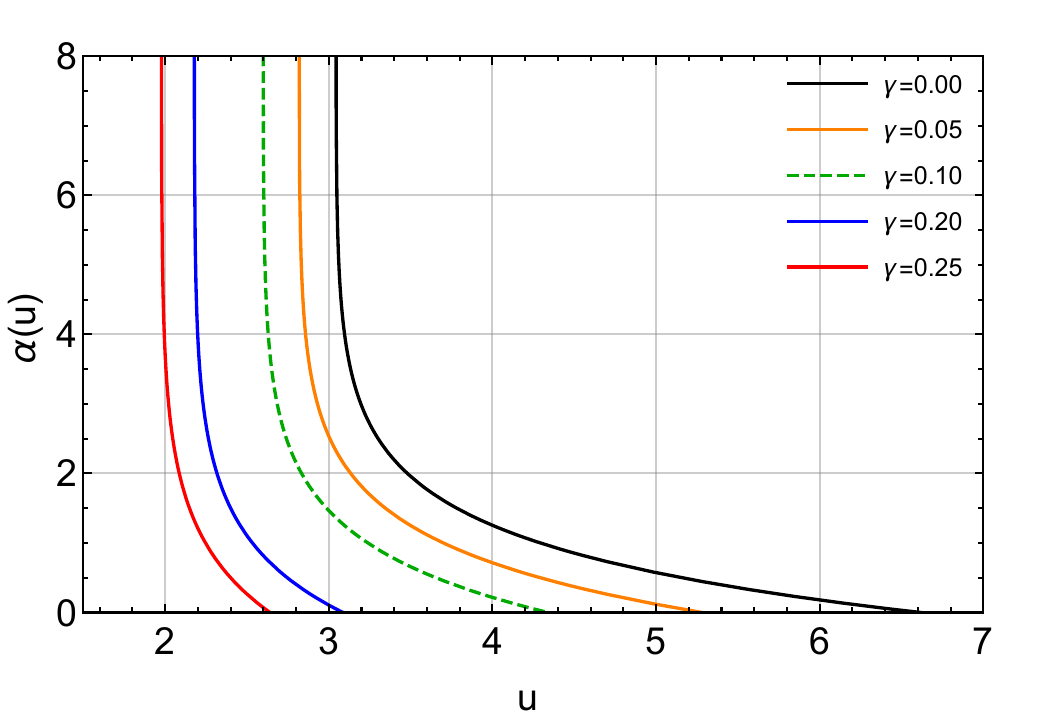}\\
\end{tabular}
\caption{The plot of deflection angle as function of impact parameter, $u$ for fixed value of global monopole term $\gamma$=0.1 (left) and LV parameter $\kappa \eta^2$=0.1 (right) respectively.Dashed green colored plot shows Schwarzschild deflection angle ($\gamma$=$\kappa \eta^2$)}.
\label{Fig:8}
\end{figure*}

Therefore, we can write  deflection angle in terms of the strong deflection limit coefficient $C^{(1)}$ and $C^{(2)}$ upto first order of impact parameter,
\begin{equation}
 \alpha^*(u)=-C^{(1)}\log (\frac{u}{u_{phs}}-1)+ C^{(2)} + \mathcal{O} (u-u_{phs})...... .
 \label{eqn:27}
\end{equation}
The variation of $C^{(1)}$ and $C^{(2)}$ versus BH parameter are shown in  Figs.~(\ref{Fig:6}) and (\ref{Fig:7})
 respectively. The first strong deflection limit coefficient decreases with the global monopole charge parameter $\kappa\eta^{2}$ and increases with the LV parameter $\gamma$, whereas the second coefficient exhibits the opposite behavior. These variations are reflected in the deflection angle, as shown in Fig.~(\ref{Fig:8}), where we plot the deflection angle as a function of the impact parameter $u$ for a fixed value of the global monopole parameter $\gamma = 0.1$ (left panel) and a fixed value of the Lorentz-violating parameter $\kappa\eta^{2} = 0.1$ (right panel). There exists a particular value where $\kappa\eta^{2} = \gamma$, for which the effects of the global monopole charge and the LV parameter cancel each other, and the BH effectively behaves like a Schwarzschild BH. This condition is indicated by the dotted curve in Fig.~(\ref{Fig:8}). Consequently, the deflection angle increases and shifts away from the BH center with increasing $\kappa\eta^{2}$, while the opposite trend is observed for increasing $\gamma$. This shift is consistent with the variation of the impact parameter shown in Fig.~(\ref{Fig:4}).
The deflection angle diverges at a particular value of the minimum impact parameter for specific values of the BH parameters. As a result, photons undergo an infinite number of loops in the photon region. These photons are detected by a distant observer in the form of circular rings, known as relativistic Einstein rings. The corresponding images are called relativistic images, a term coined by K.~S.~Virbhadra and G.~F.~R.~Ellis in 2000~\cite{Virbhadra2000}. Relativistic images are prime signatures of SGL, in contrast to the primary and secondary images formed in the weak gravitational lensing regime. In the following section, we discuss how the BH parameters affect observables in the SGL scenario.

\newpage
\section{Observables:Relativistic Einstein's Rings, Magnification, Limiting Angular Radius And Image Separation, Image Flux Ratio} \label{section4}
In this section, we investigate the Einstein ring , image magnification and separation of the BH.  Let $\beta$ and $\theta$ measures the angular position of the  source  (S) and the image (I) respectively (see Fig.~\ref{Fig:2}). These are related by the well-known Virbhadra-Ellis lens equation~\cite{Virbhadra2000},

\begin{equation}
    \boxed{ \tan{\beta}=\tan{\theta}-\frac{D_{LS}}{D_{OS}}\ [\ \tan{\theta}+\tan(\alpha-\theta)\ ]}
    \label{eqn:28}
\end{equation}
Consider Source behind the lens and Source-lens -Observer are aligned perfectly along optic axis ${\tan{\theta}}$=${\theta}$\cite{Bozza2010}.Now equation~(\ref{eqn:28}) reduces to \cite{Bozza2008},
\begin{eqnarray}
    \beta=\theta-\frac{D_{LS}}{D_{OS}}\Delta{\alpha_{n}}
    \end{eqnarray}
    Where $\Delta{\alpha_{n}}$ is the deflection angle of $n^{th}$ image. Using equation ~(\ref{eqn:27}) and~(\ref{eqn:28}) the angular position of nth relativistic Einstein image is given by \cite{Bozza2002,Eiroa2011}
  \begin{eqnarray}
    \theta_{n}=\theta^{0}_{n}+\frac{({D_{OL}+{D_{LS}}})}{D_{LS}}.\frac{u_{m}e_{m}}{D_{OL}C^{(1)}}(\beta-\theta^{0}_{n}).
    \end{eqnarray}
    Here, ${\theta}^{0}_{n}$ is some reference 
    initial position of the image and true  for both sides of lens.This expression contains strong gravitational lensing coefficient also. Using, $\beta=0$ and $D_{OS}=2\ D_{OL}$, we obtain 
    the position of the outermost relativistic Einstein ring given by,  
     \begin{eqnarray}
    \theta^{E}_{n}=\frac{u_{m}}{D_{OL}}(1+e_{n})
    \label{eqn:31}
     \end{eqnarray}
    \text{ here $e_n$ is}  
    \begin{align*}
    & e_{n}=\exp{\frac{C^{(2)}}{C^{(1)}}-\frac{2\ n\ \pi}{C^{(1)}}}.
      \end{align*}
In the limiting case $e_{n}$ vanishes as $n$ approaches infinite.The impact parameter can we written as $u_{phs}=\theta_{\infty}\ D_{OL}$.Therefore innermost relativistic Einstein ring determine the edge of black hole.

\subsection{Relativistic Einstein's Rings}
As null geodesics captured in photon region, it makes infinite loops before escaping strong gravitational field. For the single looping observer sees one bigger images than original source on both sides in the form of circular ring, called Einstein ring. X.X.Zeng et al. and A.Anand have been studied holographic Einstein ring in global monopole, the monopole parameter change the brightness of Einstein ring~\cite{Zeng2024Holographic,Anand2025}. The respective inner ring are known as relativistic Einstein ring. The angular size is estimated using equation~\eqref{eqn:31}.The following figure shows Einstein rings for two astrophysical BH $SgrA^{*}$ and $M87^{*}$ in upper and lower panel respectively.

\begin{figure*}[htbp]
\begin{tabular}{c c}
\includegraphics[width=0.45\linewidth, height=0.35\textheight]{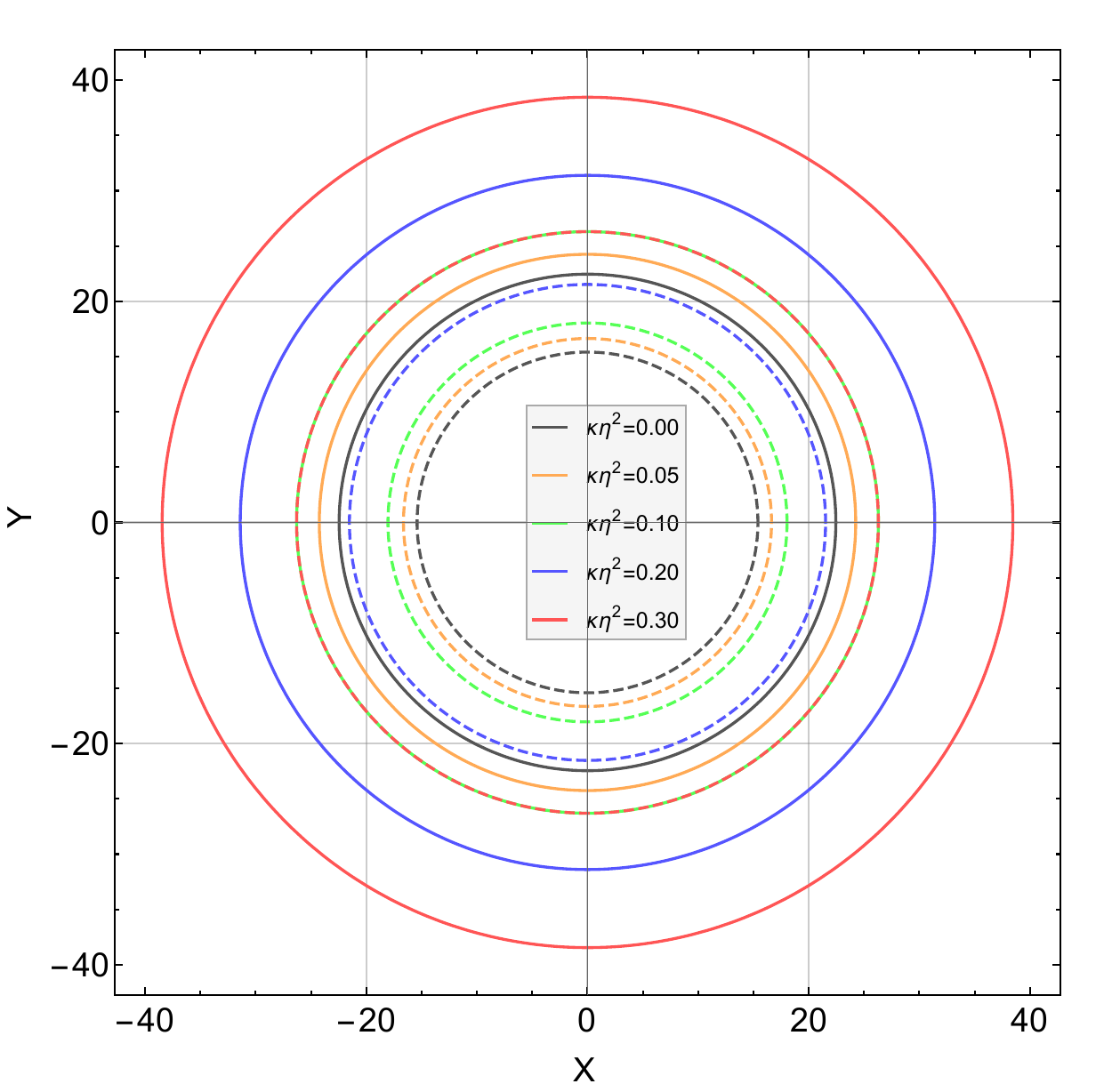}
\includegraphics[width=0.45\linewidth, height=0.35\textheight]{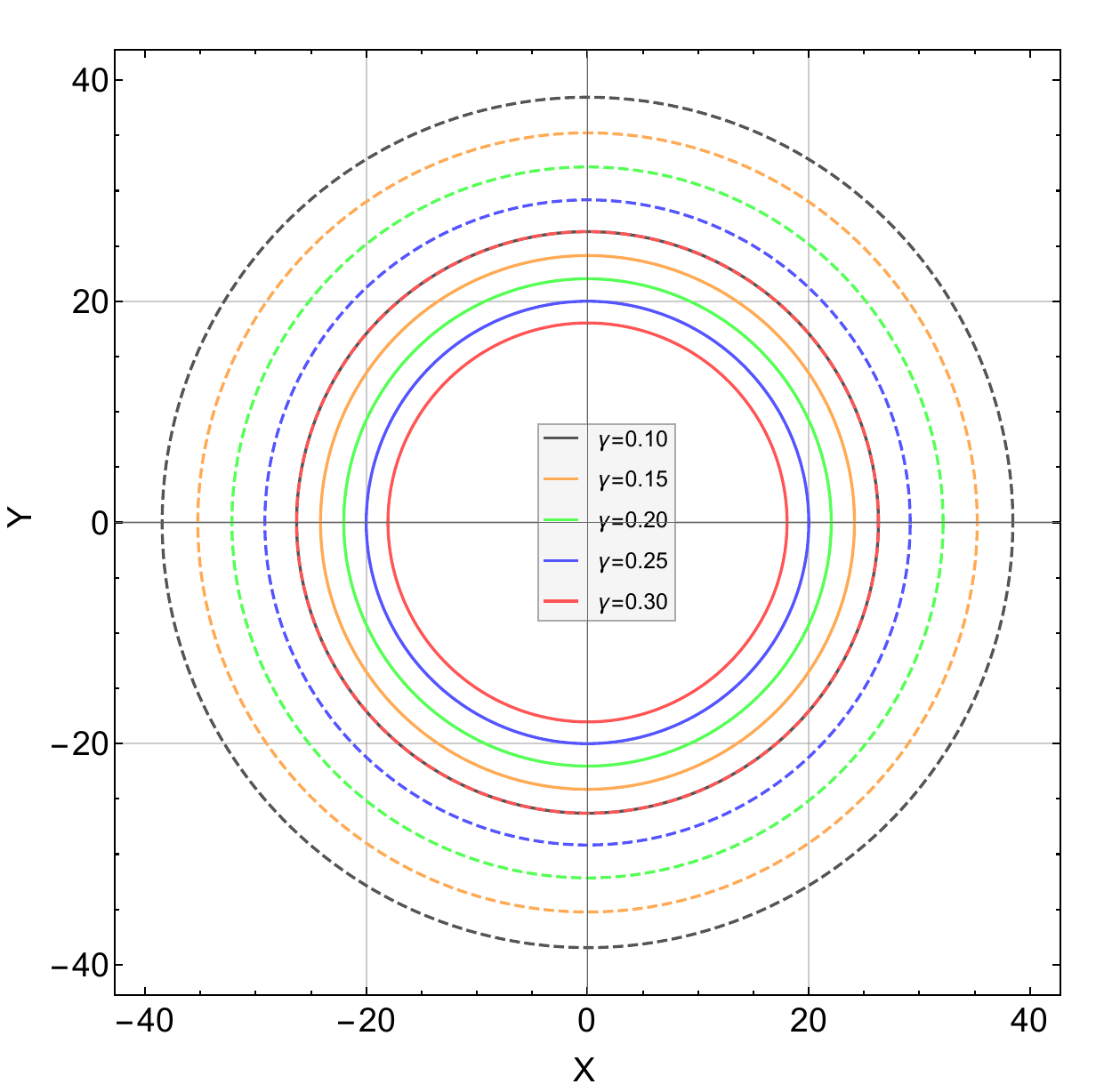}\\
\includegraphics[width=0.45\linewidth, height=0.35\textheight]{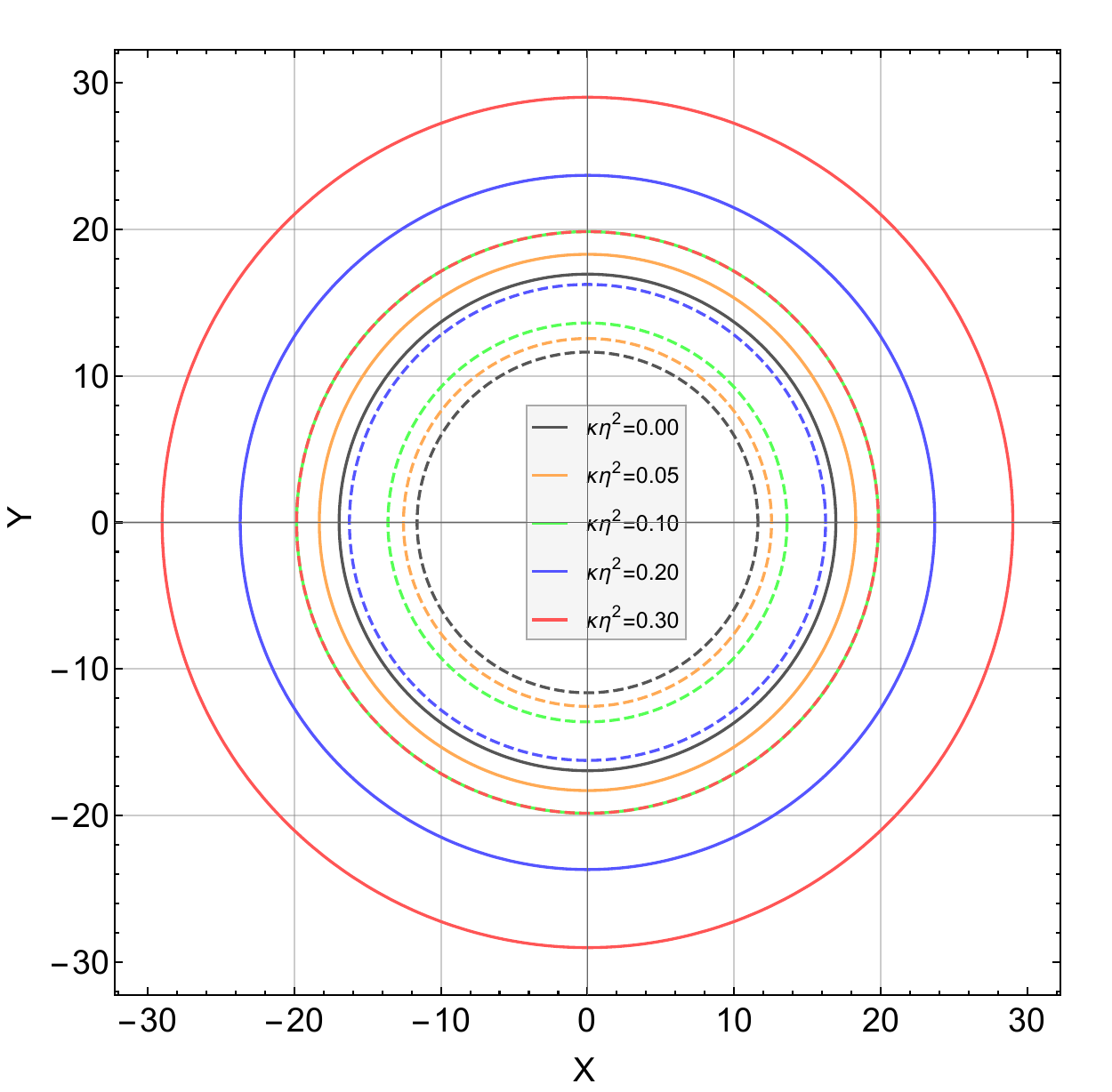}
\includegraphics[width=0.450\linewidth, height=0.35\textheight]{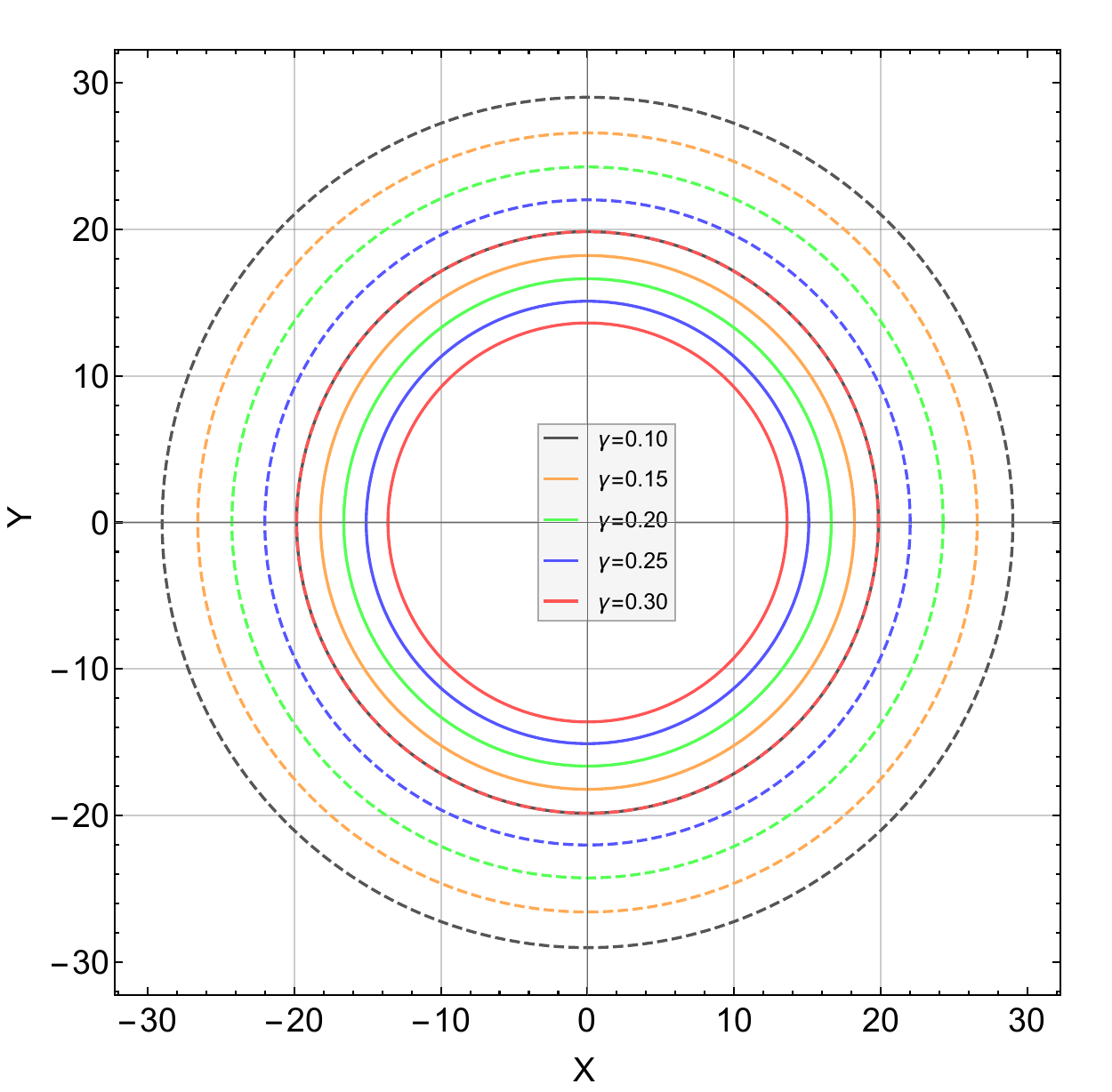}\\
 \end{tabular}
\caption{Upper panel: The plot of outermost relativistic Einstein ring (in $\mu arsec$) for fixed value of magnetic charge parameter, $\gamma=0.1$ (solid circle) and $\gamma=0.3$ (dashed circle) left and $\kappa\eta^2=0.1$ (solid circle) and $\kappa\eta^2=0.3$ (dashed circle) for $SgrA^{*}$ . Lower panel: The plot of  Einstein ring for fixed value of magnetic charge parameter, $\gamma=0.1$ (solid circle) and $\gamma=0.3$ (dashed circle) left and $\kappa\eta^2=0.1$ (solid circle) and $\kappa\eta^2=0.3$ (dashed circle) for $M87^{*}$ .}
 \label{Fig:9}
\end{figure*}
The outermost relativistic Einstein ring radius increases with the global monopole charge parameter and decreases with the LV parameter in the range $0.0$–$0.3$, as shown in Fig.~(\ref{Fig:9}). Furthermore, the Einstein ring associated with Sgr~A$^{*}$ is larger than that of M87$^{*}$, reflecting the dependence of the lensing observables on the mass–distance ratio of the lens. This behavior highlights the sensitivity of SGL signatures to modifications of gravity and provides a potential observational avenue to constrain such deviations using high-resolution astronomical observations. In the next subsection, we discuss the variation of the magnification, image separation, and flux ratio in terms of the strong lensing coefficients. These observables are sufficient to probe the nature of the BH and to distinguish between different gravity models.
\subsection{Magnification}
\begin{figure*}[htbp]
\centering
\begin{tabular}{c c}
\includegraphics[width=0.45\linewidth,height=0.4\textheight]{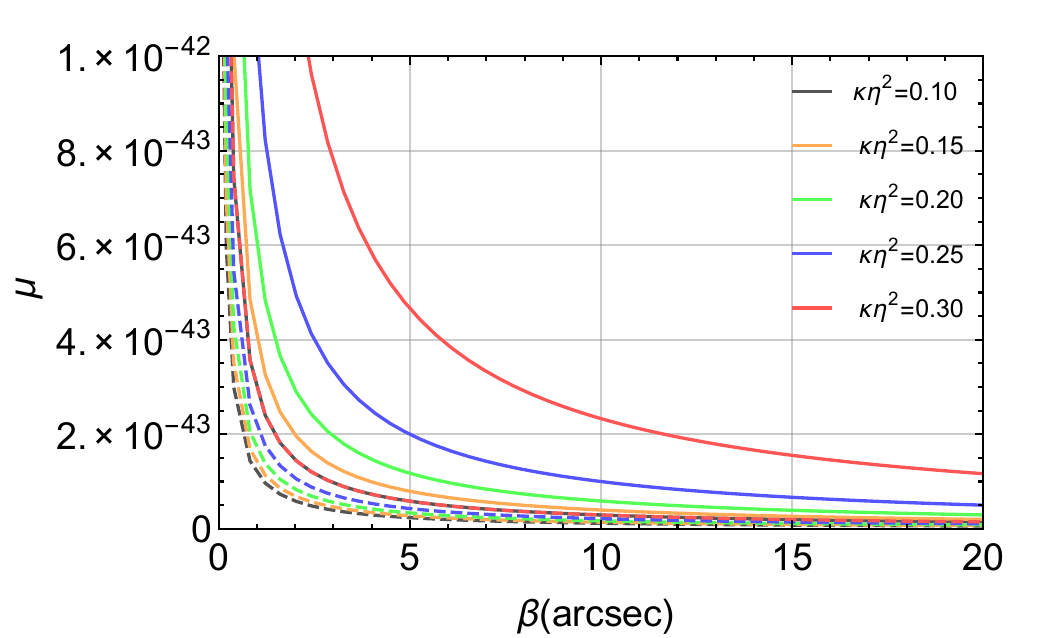}
\includegraphics[width=0.45\linewidth,height=0.4\textheight]{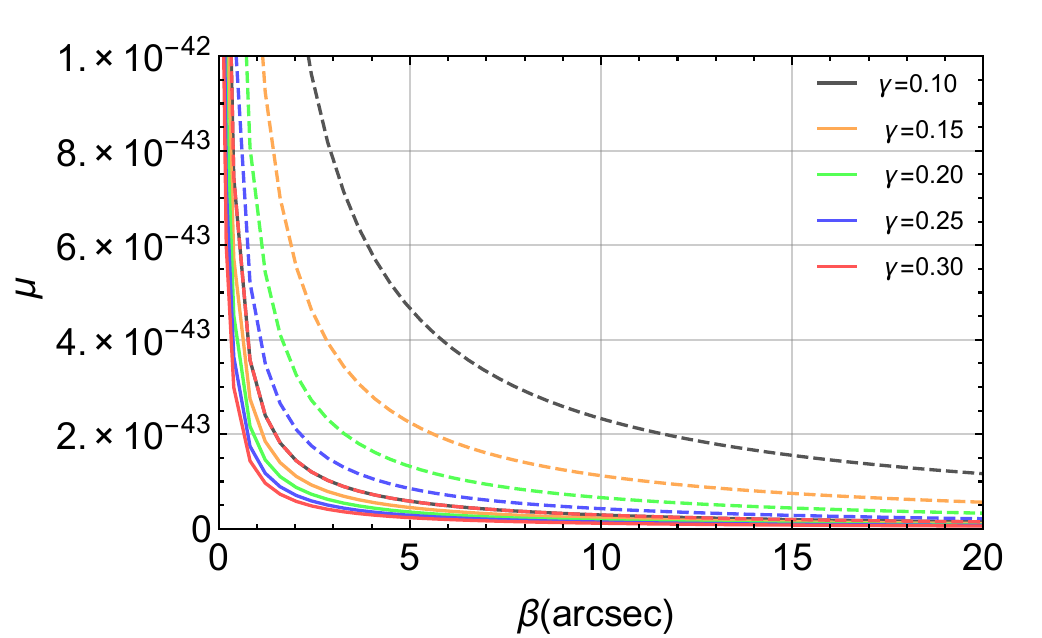}\\
\includegraphics[width=0.45\linewidth,height=0.4\textheight]{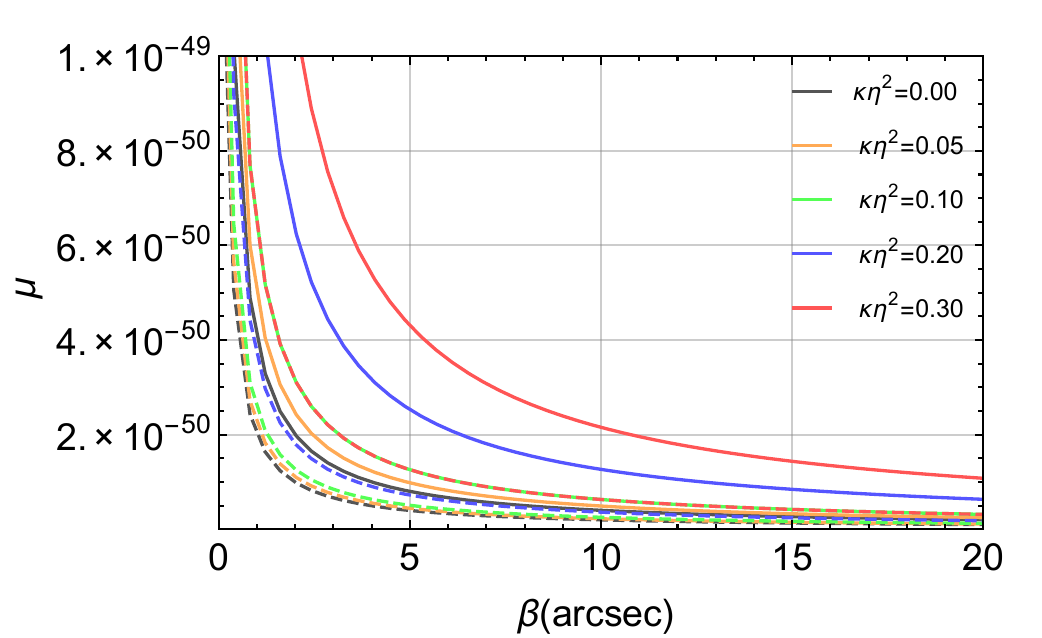}
\includegraphics[width=0.45\linewidth,height=0.4\textheight]{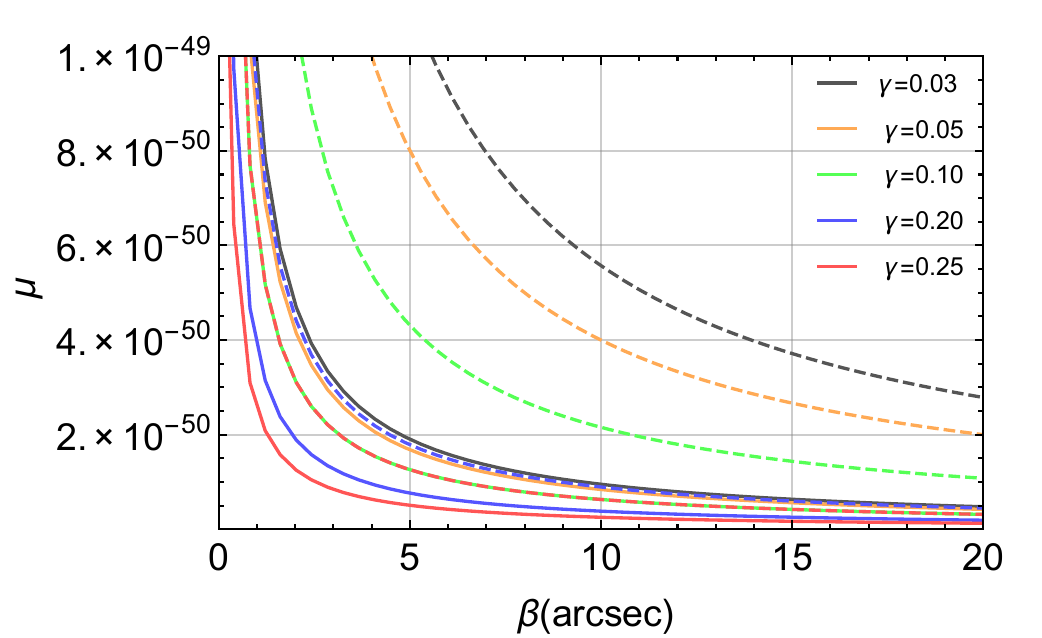}\\
\end{tabular}
\caption{Upper panel: Absolute magnification (A.M.) of the first relativistic image for fixed values of $\gamma=0.1$ (solid line) and $\gamma=0.3$ (dashed line) in the left plot, and $\kappa\eta^{2}=0.1$ (solid line) and $\kappa\eta^{2}=0.3$ (dashed line) in the right plot, as a function of the source position (in arcsec) for Sgr~A$^{*}$. Lower panel: Same as the upper panel, but for M87$^{*}$, with $\gamma=0.1,\,0.3$ (left) and $\kappa\eta^{2}=0.1,\,0.25$ (right).}
\label{Fig:10}
\end{figure*}

The surface brightness of the source does not change by gravitational lensing because energy and number of photons are unaffected in GL. The light rays are spread in large area rise to greater flux.This results in the magnification of image. The magnification of relativistic images is an important observable and it is defined as the ratio solid angles of the images and source respectively~\cite{Narayan1992}. It is determined by the following relation,
    \begin{eqnarray}
       \mu=\left(\frac{\sin\beta \ \partial\beta}{\sin\theta \ \partial\theta}\right)^{-1}.
       \end{eqnarray}
       ${\beta}$ and ${\theta}$ are the angular size of the source and the images respectively.
       The magnification in terms of strong deflection limit coefficients and the position of the relativistic image and the source is given by the following expression-, \cite{Schneider1992,Bozza2005,Perlick2004,Islam2024strong},
        \begin{eqnarray}
         \mu_{n}=\frac{1}{\beta \  C^{1}}\frac{{u_{phs}}^2}{{D_{OL}}^{2}}\ {e_{n}\ (1+e_{n}})\ (1+\frac{D_{OL}}{D_{LS}}),
         \label{am}
\end{eqnarray}
where ${\beta}$ and ${C^{1}}$ represents the source positions and first lensing coefficient. In case of perfect alignment $\beta=0$ magnification gets its maximum value. It have both positive and negative parity. We study absolute magnification (only positive value) which is observable and independent of their parity.\\
In Fig.~(\ref{Fig:10}), we plot the absolute magnification $\mu$ of the first relativistic image as a function of the source position $\beta$ for Sgr~A$^{*}$ (upper panel) and M87$^{*}$ (lower panel), for different values of the BH parameters $\kappa\eta^{2}$ and $\gamma$. The magnification increase slightly with the global monopole parameter $\kappa\eta^{2}$ and decreases with the Lorentz-violating parameter $\gamma$ for both astrophysical black hole. Moreover, the magnification is systematically larger for Sgr~A$^{*}$ than for M87$^{*}$, reflecting the stronger lensing efficiency associated with its mass–distance configuration. The shift of the magnification curve away from the source position indicates that the lensing observables are sensitive to deviations from general relativity, thereby offering a potential avenue to constrain modified gravity parameters through precision lensing observations. 

\subsection{Limiting Angular Radius and Image Separation}
\begin{figure*}[htbp]
\centering
\begin{tabular}{c c}
\includegraphics[width=0.45\linewidth,height=0.4\textheight]{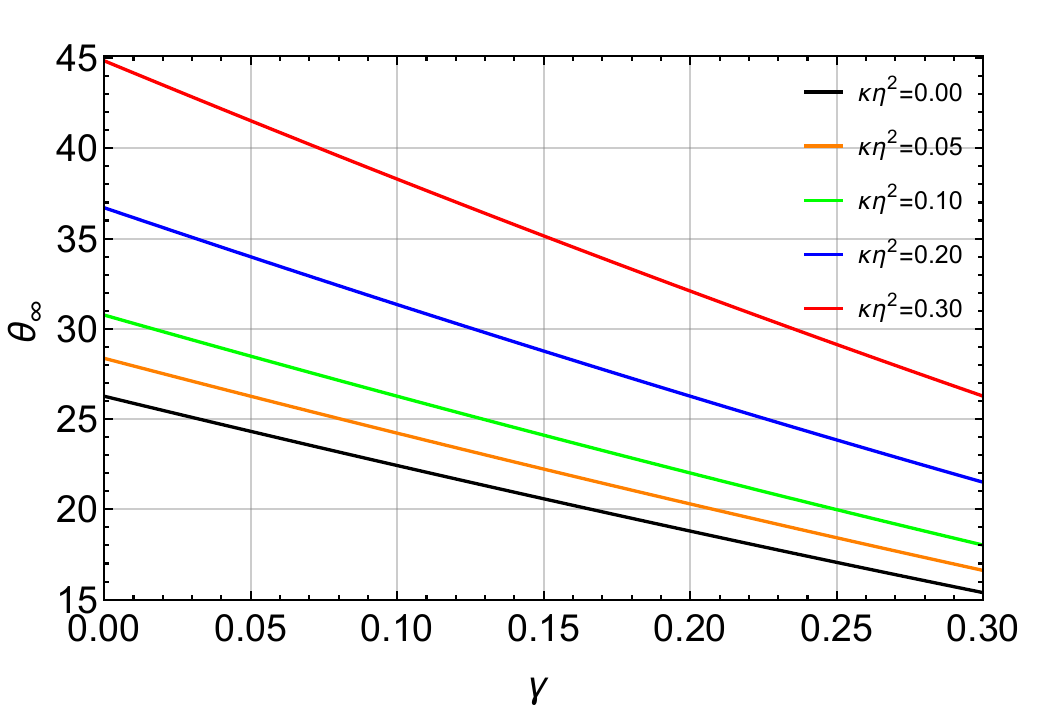}
\includegraphics[width=0.45\linewidth,height=0.4\textheight]{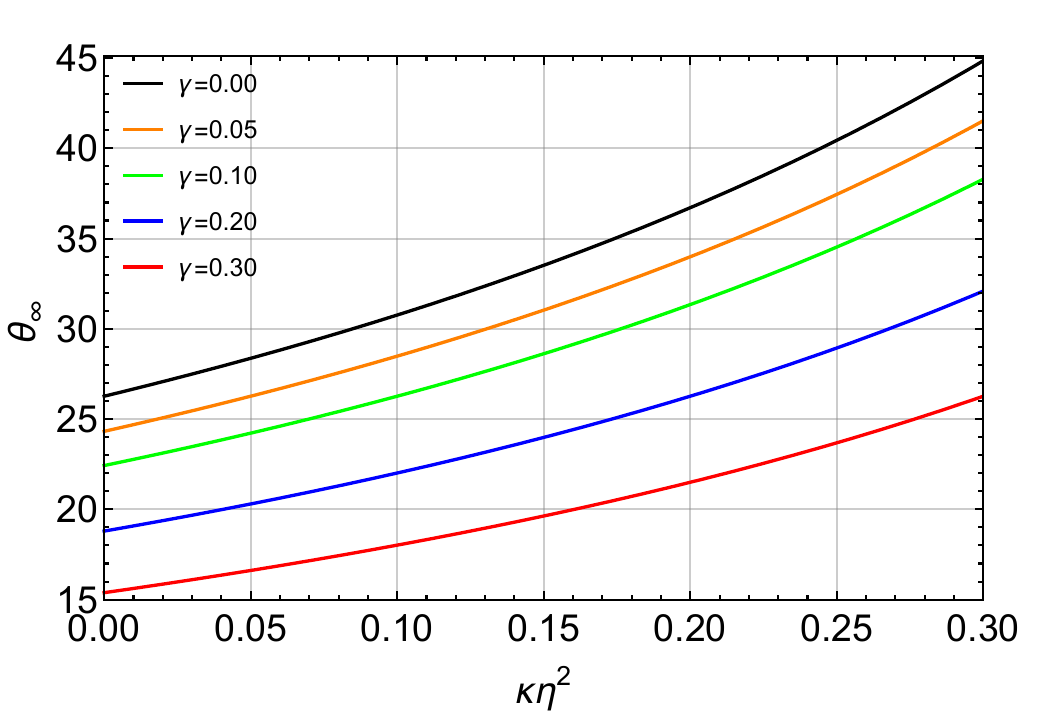}\\
\includegraphics[width=0.45\linewidth,height=0.4\textheight]{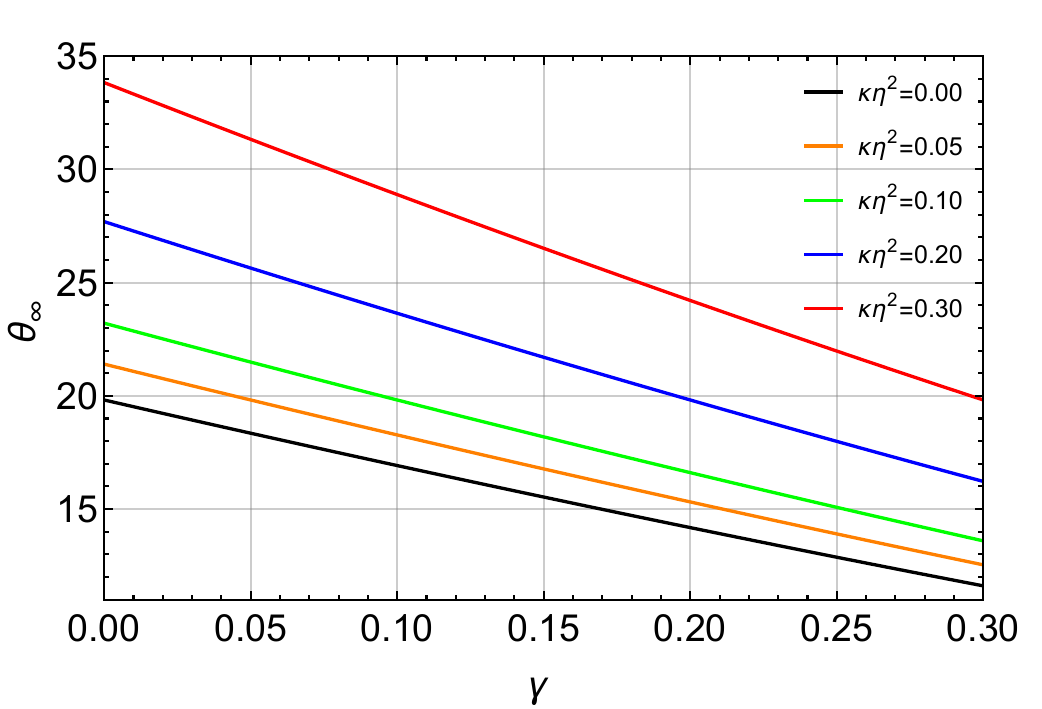}
\includegraphics[width=0.45\linewidth,height=0.4\textheight]{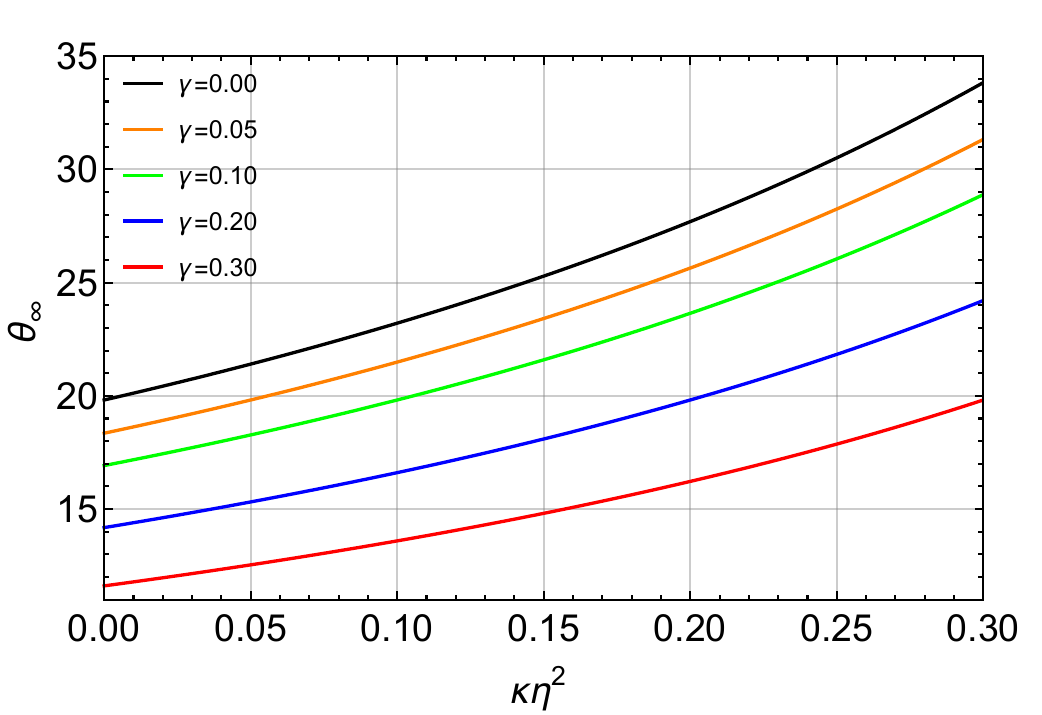}\\
\end{tabular}
\caption{Upper panel: Variation in $\theta_{\infty}$ for Sgr~A$^{*}$ in ${\mu arcsec}$ versus $\gamma$  in the left plot,  and $\kappa\eta^{2}$  in the right plot. Lower panel: Variation in $\theta_{\infty}$ for M87$^{*}$ versus $\gamma$  in the left plot,  and $\kappa\eta^{2}$  in the right plot}
\label{Fig:11}
\end{figure*}

The limiting photon radius describes accumulation of infinite relativistic images projected on the observer's sky.The variation of limiting photon angular radius shown in Fig.~(\ref{Fig:11}). It increase also with $\kappa \eta^2$ and decrease with ($\gamma$) LV parameter respectively.

The image separation of relativistic images plays an important role in determining the strong lensing coefficients.
\begin{figure*}[htbp]
\centering
\begin{tabular}{c c}
\includegraphics[width=0.45\linewidth,height=0.4\textheight]{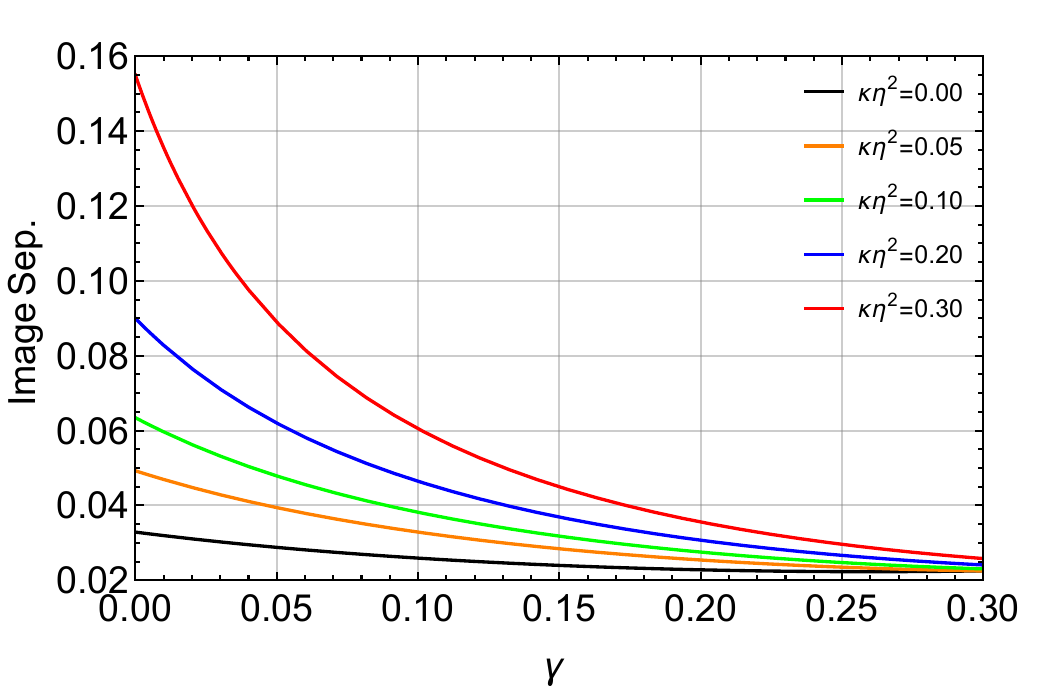}
\includegraphics[width=0.45\linewidth,height=0.4\textheight]{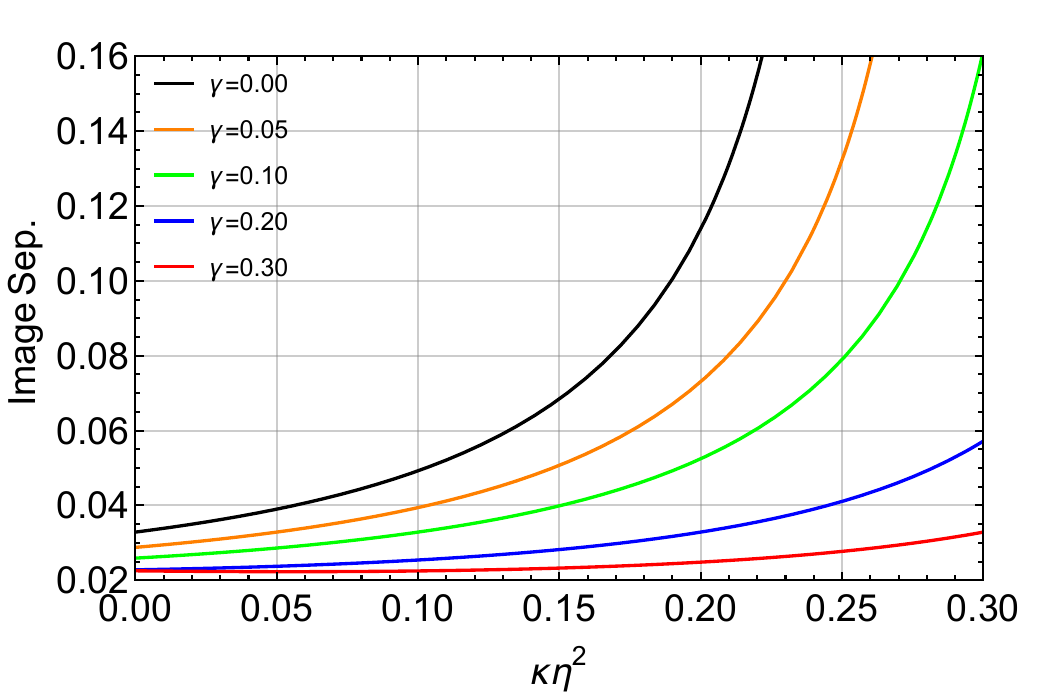}\\
\includegraphics[width=0.45\linewidth,height=0.4\textheight]{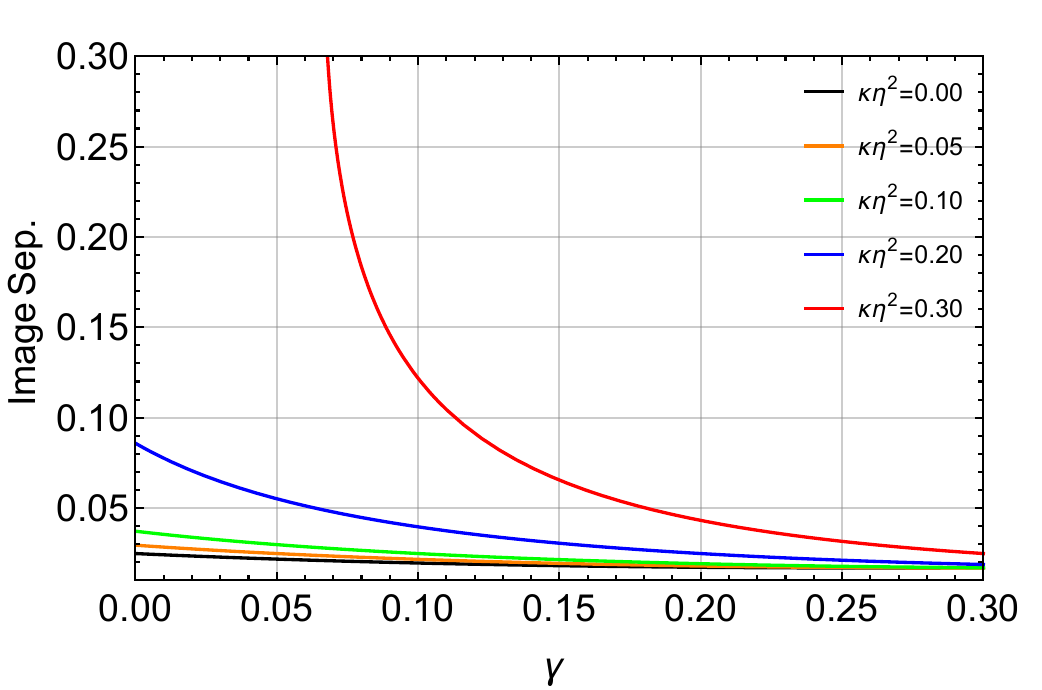}
\includegraphics[width=0.45\linewidth,height=0.4\textheight]{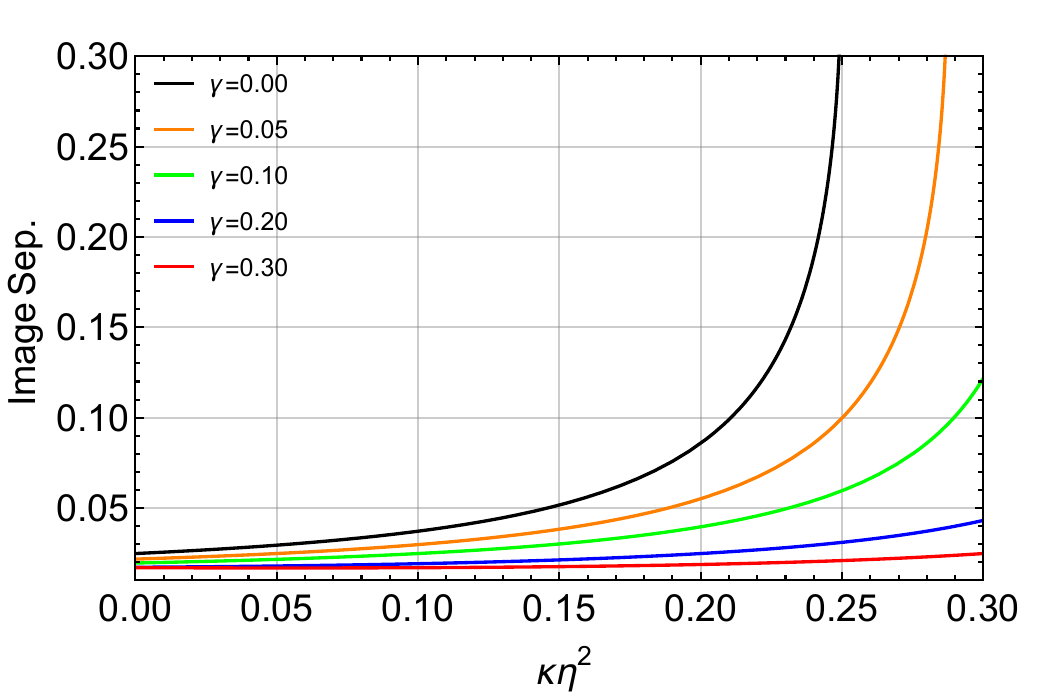}\\
\end{tabular}
\caption{Upper panel: Separation of the first relativistic image (in $\mu arsec$) as a function of the source position (in arcsec) for Sgr~A$^{*}$, corresponding to fixed values of $\gamma=0.1$ (solid line) and $\gamma=0.3$ (dashed line) in the left plot, and $\kappa\eta^{2}=0.1$ (solid line) and $\kappa\eta^{2}=0.3$ (dashed line) in the right plot. Lower panel: Same as the upper panel, but for M87$^{*}$, with $\gamma=0.1,\,0.3$ (left) and $\kappa\eta^{2}=0.1,\,0.25$ (right).}
\label{Fig:12}
\end{figure*}
Therefore the outermost images is resolved while others bunch of images remained packed at $\theta_{\infty}$. The separation between the first image and the packed set of images is defined as $S = \theta_1 - \theta_{\infty}$. Considering two astrophysical BHs as gravitational lenses, the image separations are shown in Fig.~\ref{Fig:12}. It is evident that the separation increases with the global monopole charge $\kappa\eta^2$ but decreases with the Lorentz-violating parameter $\gamma$. Moreover, the image separation is larger for Sgr~A$^{*}$ than for M87$^{*}$. 
\subsection{Image Flux Ratio}
It is defined as ratio of flux of first image to flux accumulated by rest of the images.Image flux ratio is given by,
\begin{equation}
    r=\frac{\mu_{1}}{\sum_{n=2}^{\infty} \mu_n}.
\end{equation}
In terms of SGL coefficient this sum can be expressed as~\cite{Bozza2002,Chakraborty2017},

\begin{figure*}[htbp]
\centering
\begin{tabular}{c c}
\includegraphics[width=0.45\linewidth, height=0.4\textheight]{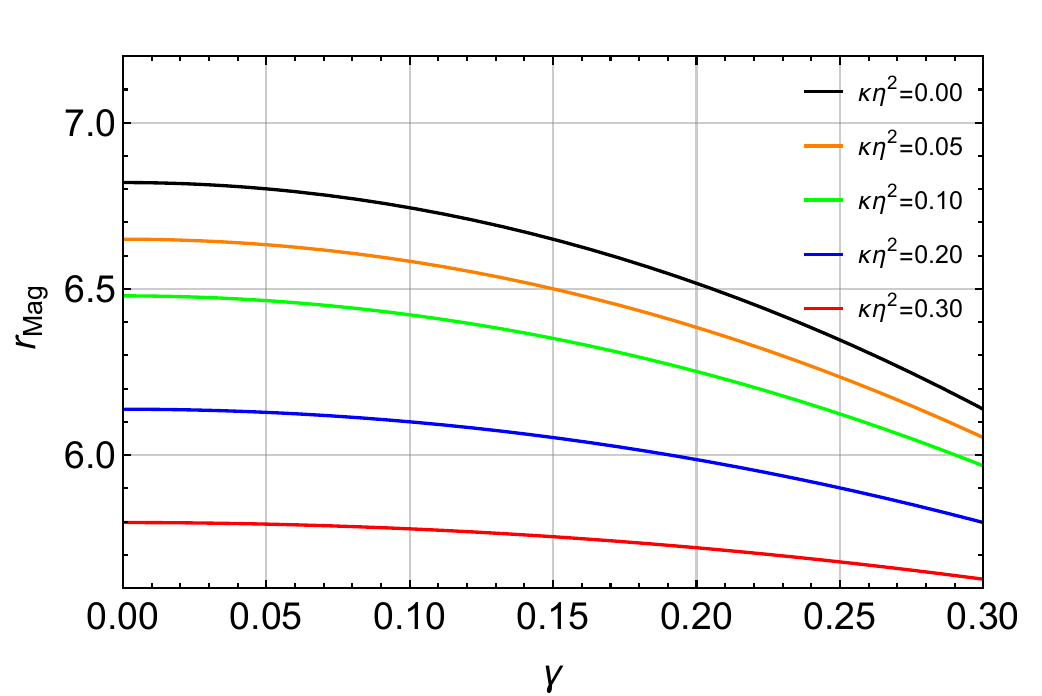}
&
\includegraphics[width=0.45\linewidth, height=0.4\textheight]{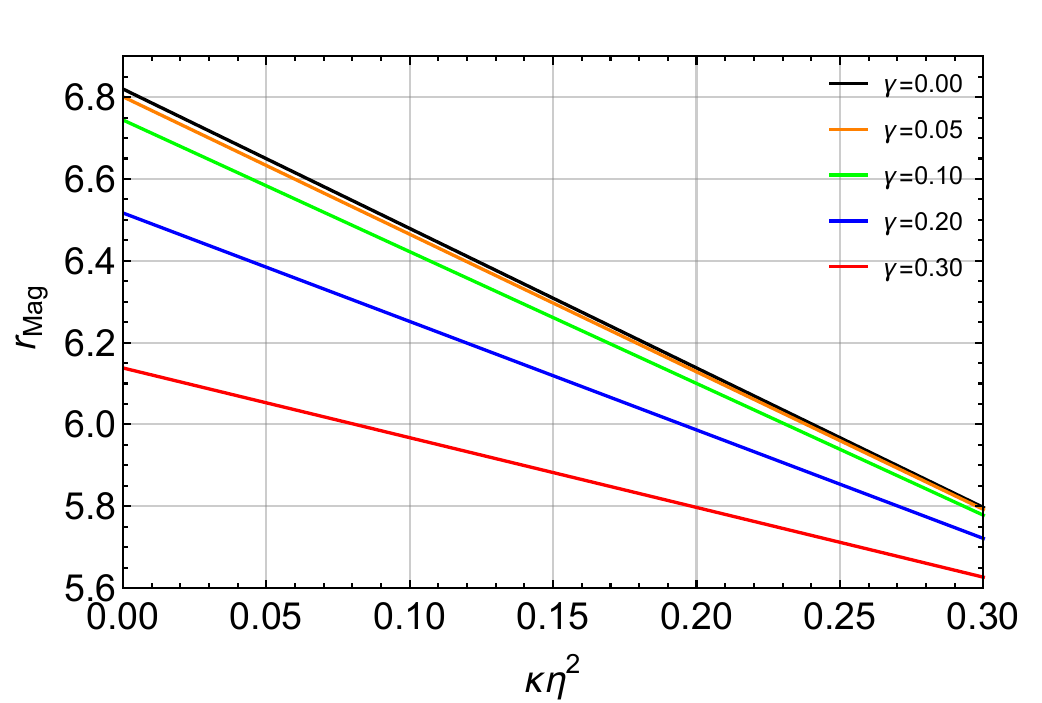}
\end{tabular}
\caption{Observables $r_{\mathrm{Mag}}$ versus $\gamma$ and $\kappa\eta^2$, respectively.}
\label{Fig:13}
\end{figure*}
\begin{equation}
    \sum_{n=2}^{\infty} \mu_n = \frac{u^2_{m}D_{OS}} {C^{(1)}\ \beta\ D_{OL}^2D_{LS}} \exp{\left(\frac{C^{(2)}-2\ n\ \pi}{C^{(1)}}\right)} \left[1+\exp{\left(\frac{C^{(2)}-2\ n\ \pi}{C^{(1)}}\right)} \right],
\end{equation}
using $ \exp{\left(\frac{2 \pi}{C^{(1)}}>>1\right)}, \exp{\left(\frac{C^{(2)}}{C^{(1)}}>>1\right)}$
\begin{eqnarray}
    r=\exp{\left(\frac{2 \pi}{C^{(1)}}\right)},\\
     r_{Mag}=\frac{5\ \pi}{C^{(1)} \ ln 10}.
     \label{eqn:rmag}
\end{eqnarray}
The equation~\eqref{eqn:rmag} represents the relative magnification of the first image with respect to the other relativistic images.In terms of the above observables, the strong deflection lensing coefficients are obtained as follows.
\begin{eqnarray}
    C^{(1)}= \frac{2\ \pi}{\log r},\\
     C^{(2)}= C^{(1)}\ \log \frac{r S} {\theta_{\infty}},
\end{eqnarray}

The magnitude of the flux ratio is shown in Fig.~(\ref{Fig:13}). This ratio decreases with increasing global monopole charge and the LV parameter.
Therefore, the SGL coefficients can be estimated in terms of lensing observables. It is interesting to note that, using observational data of only the angular separation and flux ratio, one can probe the extreme regime of a strong gravitational field and identify the nature of the BH in a given gravity theory. In the next section, we discuss the BH shadow and the influence of BH parameters on it.
\section{Optical Shadow of Black Hole with Global Monopole} \label{section5}
Consider radius of photon sphere (angular position of the images${n \to \infty}$) is denoted by $\theta_{\infty}$. Therefore
angular diameter of BH shadow in terms of angular radius of photon sphere $\theta_{\infty}$ is given by $\theta_{Shadow}=2\ \theta_{\infty}$~\cite{Bozza2002,CunninghamBardeen1973,Takahashi2004}.

\begin{figure*}[htbp]
\centering
\begin{tabular}{c c}
\includegraphics[width=0.45\linewidth, height=0.35\textheight]{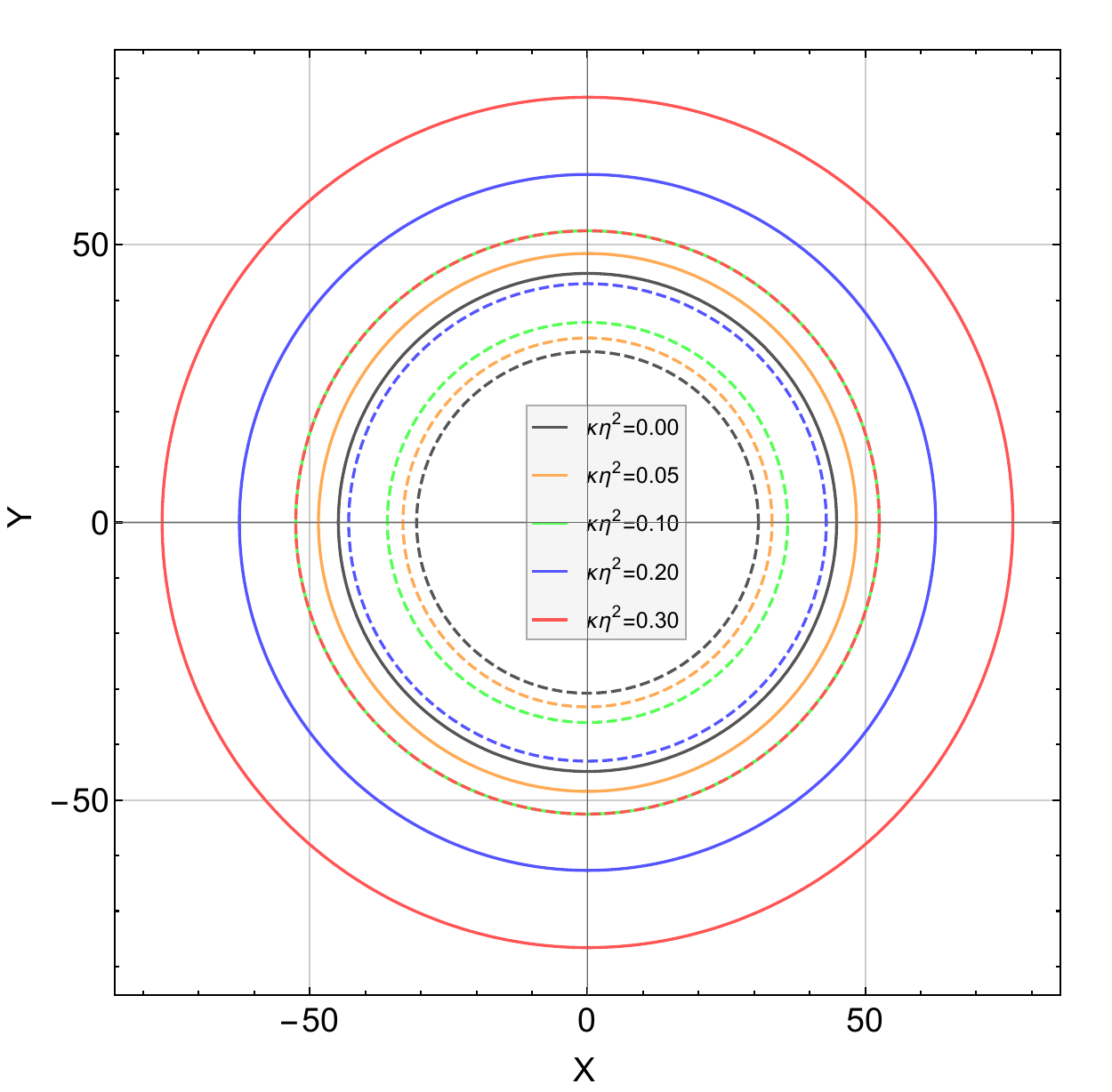}
&
\includegraphics[width=0.45\linewidth, height=0.35\textheight]{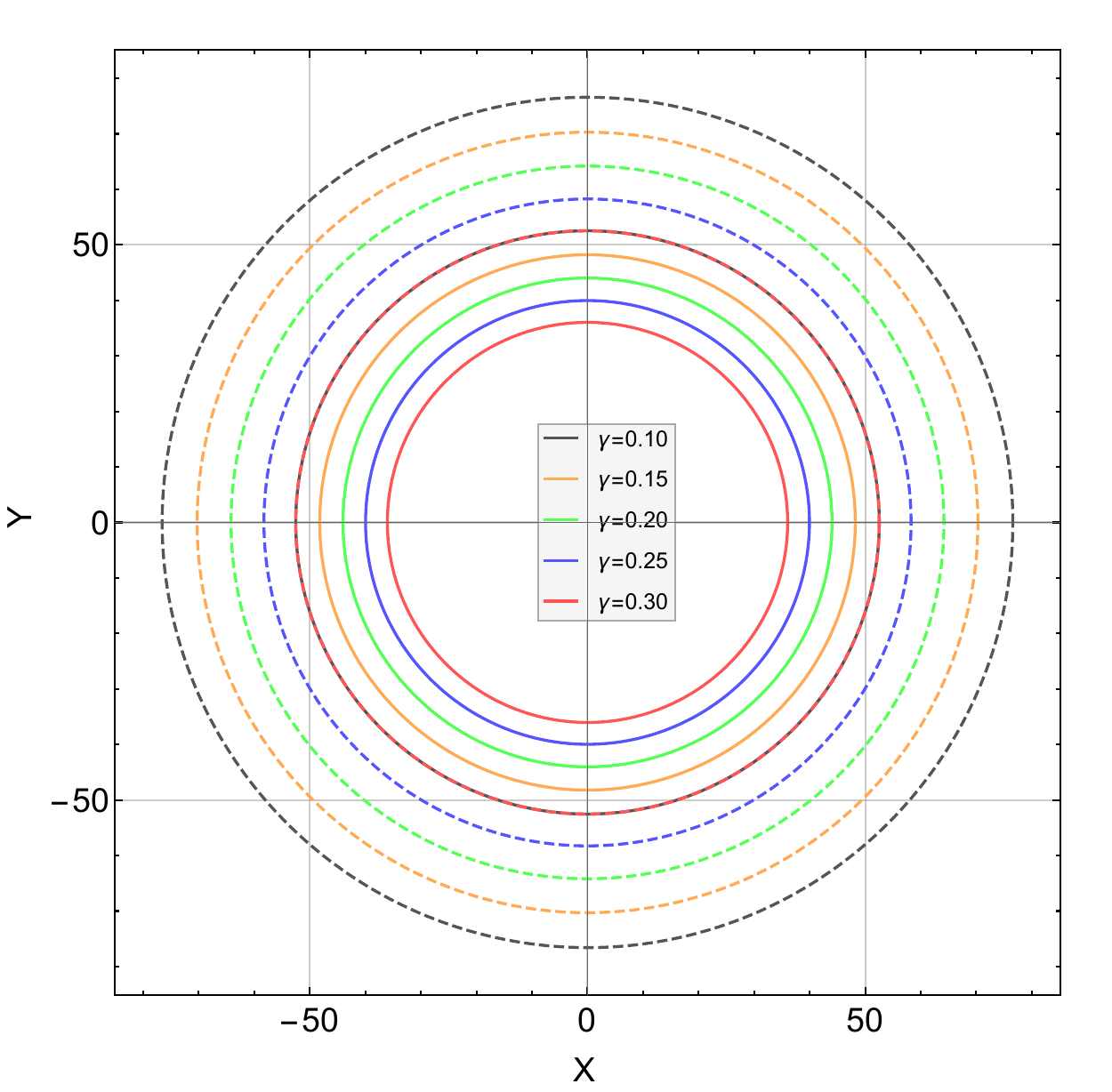}
\\
\includegraphics[width=0.45\linewidth, height=0.35\textheight]{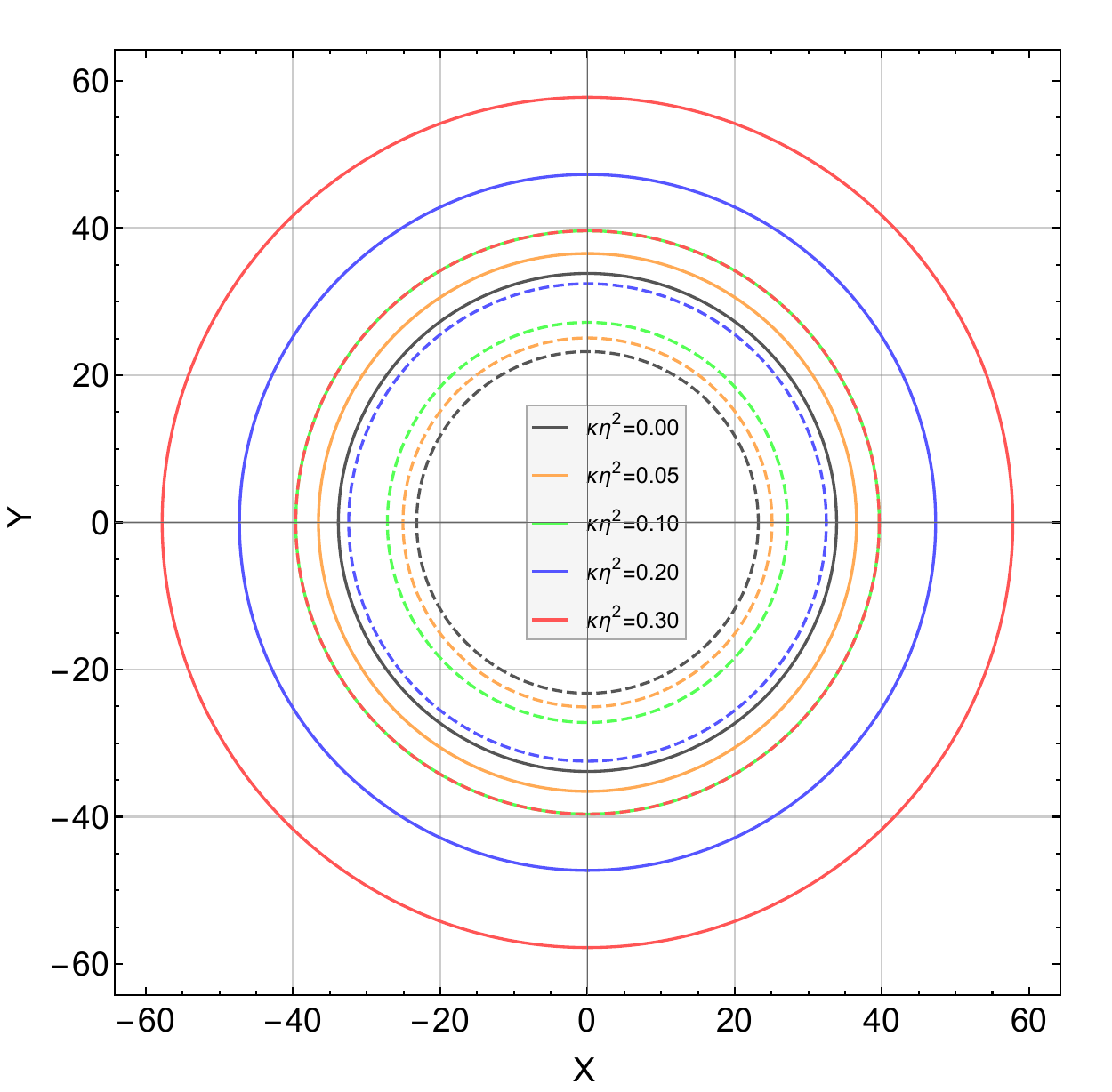}
&
\includegraphics[width=0.45\linewidth, height=0.35\textheight]{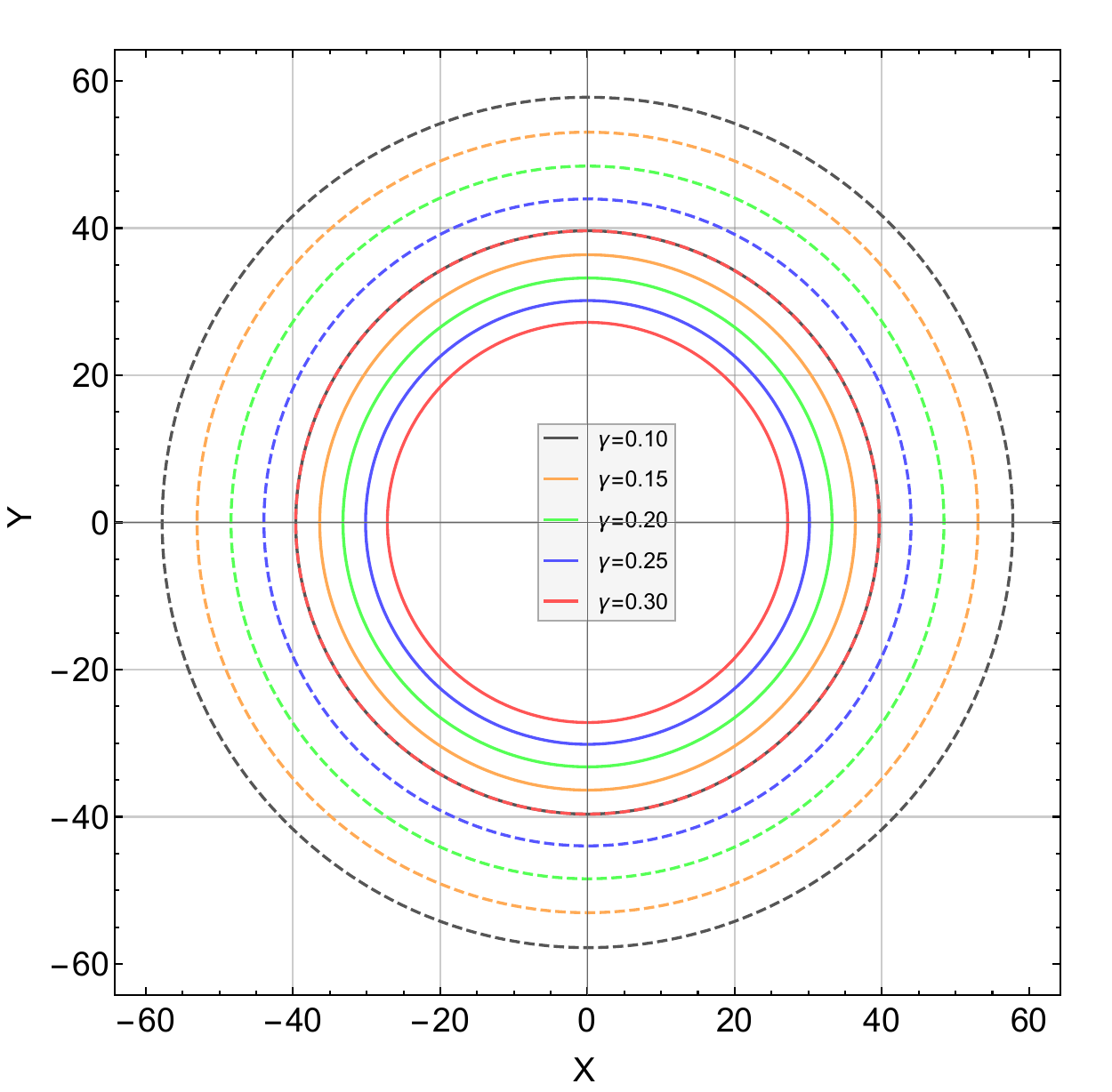}
\\
\end{tabular}
\caption{Upper panel: BH shadow plots (in $\mu arsec$) for fixed values of the LV parameter $\gamma = 0.1$ (solid circle) and $\gamma = 0.3$ (dashed circle) on the left, and the global monopole parameter $\kappa\eta^{2} = 0.1$ (solid circle) and $\kappa\eta^{2} = 0.3$ (dashed circle) on the right, for $SgrA^{*}$. 
Lower panel: Same as the upper panel, but for $M87^{*}$.}
\label{Fig:14}
\end{figure*}
The shadow image of Schwarzschild BH $(39.6192)$ lies in between the range $(30.7-76.75)$ $\mu as$ (upper panel), $(20.2-58.50)$ $\mu as$ (lower panel), estimated in Fig.~(\ref{Fig:14}). In the next section we determine the shadow in infalling gas assuming photon ring thin accretion disk model. The parameter of BH influence the impact parameter of photon ring and particle in ISCO in accretion disk.

\newpage
\section{Impact of Infalling Accretion Gas on Shadow Images} \label{section6}
In this section, we study the impact of radially infalling accreting gas on BH shadow images within a generalized framework, with the aim of analyzing the effects of the model parameters. To model the appearance of the BH surrounded by accreting matter, we
consider an optically thin, radially infalling gas and compute the radiation
detected by a distant observer located in the image plane $(X,Y)$. The observed
specific intensity at frequency $\nu_{\text{obs}}$ is given by~\cite{Jaroszynski1997bw},
\begin{equation}
    I_{\text{obs}}(\nu_{\text{obs}},X,Y)
    = \int_{\gamma} g^{3}\, j_{\nu_{\text{em}}} \, d\ell_{\text{prop}},
\end{equation}
where $g=\nu_{\text{obs}}/\nu_{\text{em}}$ is the redshift factor,
$j_{\nu_{\text{em}}}$ is the emissivity measured in the emitter's rest frame, and
$d\ell_{\text{prop}}$ is the proper length element along the photon trajectory
$\gamma$.

For a photon with four-momentum $p^{\mu}$, the redshift factor becomes
\begin{equation}
    g = \frac{p_{\mu} u_{\text{obs}}^{\mu}}{p_{\nu} u_{\text{em}}^{\nu}}
      = \frac{-p_{t}}{-p_{t}u^{t}_{\text{em}} + p_{r}u^{r}_{\text{em}}},
\end{equation}
with $u_{\text{obs}}^{\mu}=(1,0,0,0)$ for a static observer at infinity.
Assuming that the accretion flow undergoes radial free fall in the equatorial plane
$(\theta=\pi/2)$, the emitter's four-velocity is
\begin{align}
    u^{t}_{\text{em}} &= \frac{E}{|g_{tt}(r)|}, \\
    u^{r}_{\text{em}} &= -\sqrt{E^{2} - |g_{tt}(r)|},
\end{align}
where the metric function $g_{tt}(r) = -f(r;\,\kappa\eta^{2},\gamma)$ depends on
the global monopole term $\kappa\eta^{2}$ and the Lorentz-symmetry-breaking
parameter $\gamma$ through the modified lapse function of the spacetime.

For equatorial photons with impact parameter $b = L/E$, the momentum components are
\begin{align}
    p_{t} &= -E, \\
    p_{\phi} &= bE, \\
    p_{r} &= \pm E\sqrt{1 - \frac{b^{2} f(r;\,\kappa\eta^{2},\gamma)}{r^{2}}},
\end{align}
where the turning point of the radial motion determines the photon sphere and the
corresponding critical impact parameter $b_{\text{ph}}$.

The accretion flow emissivity is assumed to be monochromatic and radially decaying:
\begin{equation}
    j_{\nu_{\text{em}}}(r)
    = j_{0}\,\delta(\nu_{\text{em}} - \nu_{0})\, r^{-2},
\end{equation}

    \begin{figure}[htbp]
	\begin{center}
		{\includegraphics[width=0.8\textwidth]{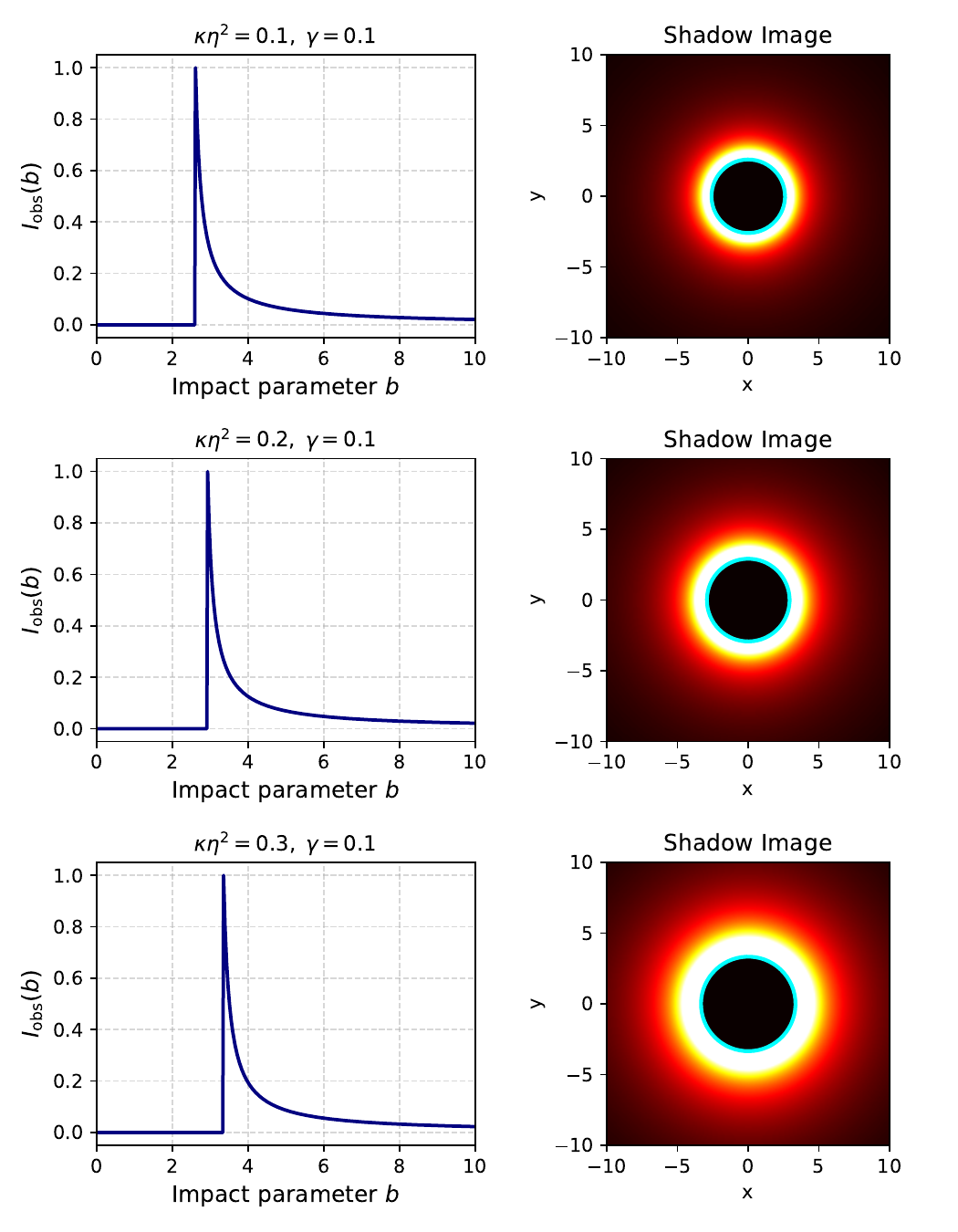}}
	\end{center}
	\caption{The left column shows the observed intensity distribution as a function of the impact parameter, while the right column presents the corresponding images of the optically thin emission region surrounding the BH.  X and Y represents the
angular celestial coordinates in the observer’s sky. Here, the value of $\gamma$ is fixed, while the parameter $\kappa\eta^{2}$ is varied.
} \label{Fig:15}
\end{figure}
    \begin{figure}[htbp]
	\begin{center}
		{\includegraphics[width=0.8\textwidth]{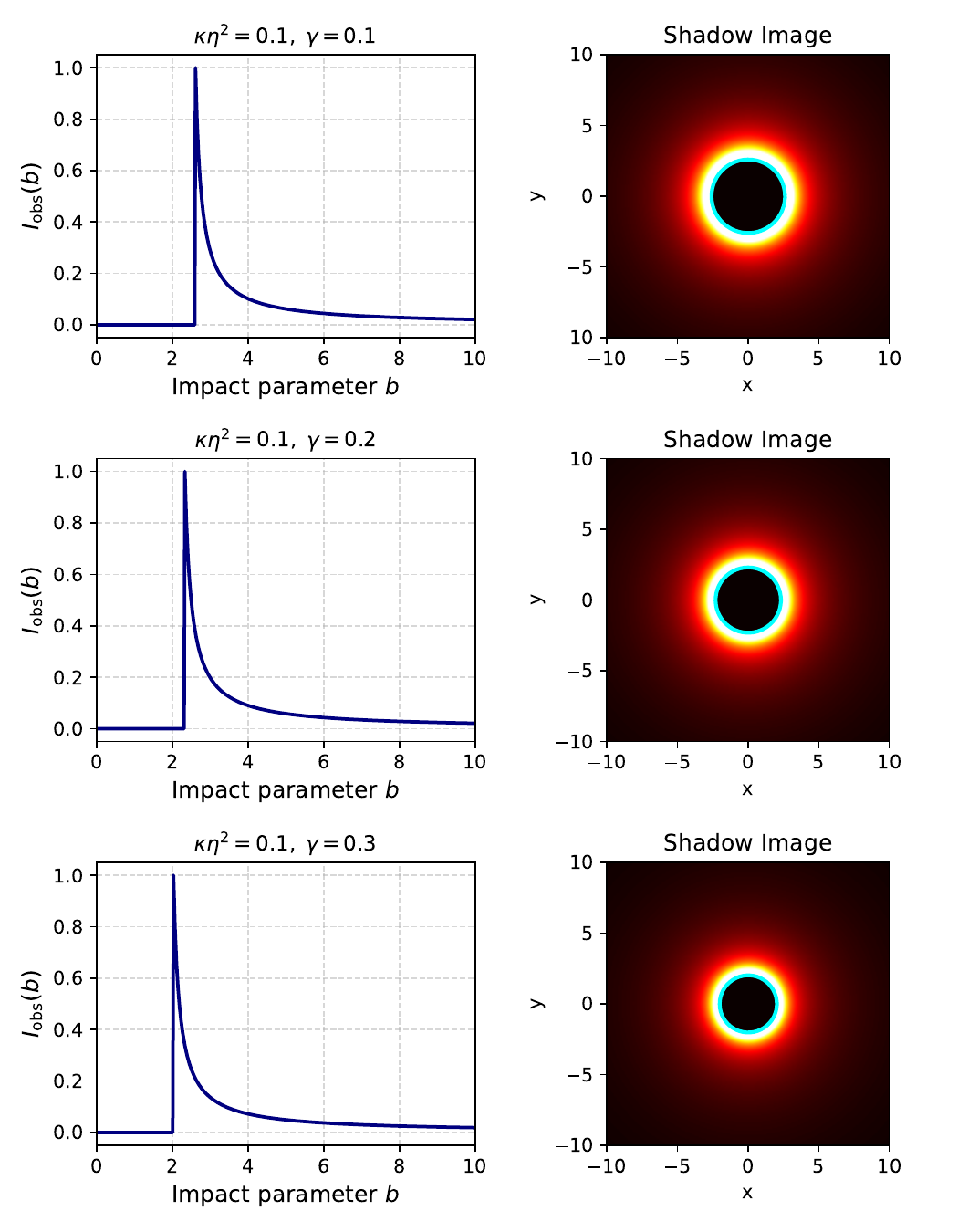}}
	\end{center}
	\caption{The left column shows the observed intensity distribution as a function of the impact parameter, while the right column presents the corresponding images of the optically thin emission region surrounding the BH.  X and Y represents the
angular celestial coordinates in the observer’s sky. Here, the value of $\kappa\eta^{2}$ is fixed, while the parameter  $\gamma$ is varied.
} \label{Fig:16}
\end{figure}

representing optically thin radial infall. The proper length element along the
photon trajectory simplifies to
\begin{equation}
    d\ell_{\text{prop}}
    = \frac{p_{t}}{p_{r}} \sqrt{g_{rr}(r)}\, dr
    = \frac{E}{|p_{r}|} \sqrt{g_{rr}(r)}\, dr.
\end{equation}

Integrating over all emitted frequencies yields the total observed intensity~\cite{Kala2020prt,Kala2024fvg,Kala2025xnb,Kukreti2025rzn},
\begin{equation}
    I_{\text{obs}}(X,Y)
    \propto \int_{\gamma}
    \frac{g^{3} p_{t}}{r^{2} |p_{r}|}\,
    \sqrt{g_{rr}(r)}\,
    dr .
\end{equation}
This expression forms the basis for constructing the BH shadow and the
associated brightness distribution. Substituting the explicit function
$f(r;\,\kappa\eta^{2},\gamma)$ allows us to quantify the influence of the global
monopole term $\kappa\eta^{2}$ and the Lorentz-symmetry--breaking parameter
$\gamma$ on the photon sphere, the critical impact parameter $b_{\text{ph}}$, and
the detailed morphology of the resulting shadow image.
    
We compute the shadow images and corresponding intensity profiles for the BH metric characterized by the global monopole term $\kappa\eta^{2}$ and the Lorentz symmetry breaking parameter $\gamma$. These parameters modify the curvature near the photon sphere, shifting the critical impact parameter $b_{\rm ph}$ and altering the brightness distribution. Ray-traced images of a radially infalling, optically thin accretion flow indicate that increasing $\kappa\eta^{2}$ leads to an enlargement of the photon-ring radius, accompanied by an enhancement of the accretion-induced emission around the shadow, as shown in Fig.~(\ref{Fig:14}), where as increasing $\gamma$ results in a reduction of the photon-sphere radius and a corresponding suppression of the emission near the shadow boundary, as illustrated in Fig.~(\ref{Fig:15}).The change in shifts of impact parameter $2.5-3.8$ are correspond to the photon impact parameter that estimated in Fig.~(\ref{Fig:4}) at fixed $\gamma= 0.1$. While it decreases slightly at fixed $\kappa \eta^2= 0.1$ form $2.5-2.0$. Evidently global monopole term and LV parameter influence shadow size the gas falling accretion disk. Qualitatively we find that BH shadow increases and decreases with BH parameter as in Fig.~(\ref{Fig:14}).  

\newpage
\section{Conclusions}
\label{section7}
In this paper, we analyzed SGL by a static BH sourced by a global monopole charge in the presence of a LV parameter. We found that the event horizon radius increases with the global monopole parameter $\kappa\eta^{2}$ and decreases with the LV parameter $\gamma$. The effective potential increases with the impact parameter for fixed values of $\kappa\eta^{2}$ and $\gamma = 0.1$, whereas it decreases with increasing global monopole charge for fixed $\gamma = 0.1$ and $u = 2.608$. The photon sphere radius and the corresponding critical impact parameter increase with the global monopole parameter and decrease with the LV parameter, respectively. This opposite behavior originates from the distinct physical roles of the global monopole charge and the LV parameter in modifying the spacetime geometry. The deflection angle shifts toward larger impact parameters with increasing $\kappa\eta^{2}$ and toward smaller impact parameters with increasing $\gamma$, demonstrating that BH parameters significantly affect the gravitational field.

The Einstein ring radius, magnification, and angular separation between images increase with the global monopole parameter and decrease with the LV parameter for both astrophysical BHs, $SgrA^{*}$ and $M87^{*}$. In addition, the BH shadow size enlarges with increasing $\kappa\eta^{2}$ and shrinks with increasing $\gamma$, while the flux ratio exhibits an opposite trend. The variation in the angular diameter of the BH shadow directly follows the behavior of the photon sphere radius. Furthermore, within the optically thin accretion disk model, we find that the BH shadow is shifted toward larger (smaller) impact parameters for fixed $\gamma = 0.1$ ($\kappa\eta^{2} = 0.1$), respectively. These results confirm that the shadow size shows consistent increasing and decreasing trends in both vacuum and thin-disk models, governed by the global monopole charge and the LV parameter.

It would be natural to extend this analysis to rotating BHs ~\cite{IyerHansen2009,Liu2025} in future work. Moreover, the deflection of massive particles in the presence of a global monopole may further enrich the study of SGL ~\cite{Jusufi2017,Rahaman2021}. With the advent of more sensitive observational facilities, such shifted relativistic images and related observables may become accessible to experiments.

\section*{Acknowledgment(s)}  
  BKV acknowledges the financial support from UGC fellowship. SK sincerely acknowledges IMSc for providing exceptional research facilities and a conducive environment that facilitated his work as an Institute Postdoctoral Fellow. We also acknowledge useful discussions with Prof. Sanjay Siwach, which significantly helped shape the development and direction of this work.

\bibliographystyle{apsrev4-2}
\bibliography{main}
\end{document}